\documentclass[onecolumn]{aastex62}

\pdfoutput=1

\usepackage{multirow}
\usepackage{graphicx}
\usepackage[caption=false]{subfig}
\usepackage{amsmath}

\submitjournal{ApJ}

\shorttitle{Star Formation in the SMOG.}
\shortauthors{Winston et al.}

\begin{document}

\title{ A  Census of Star Formation in the Outer Galaxy: the SMOG field.}

\correspondingauthor{Elaine Winston}
\email{elaine.winston@cfa.harvard.edu}

\author[0000-0001-9065-6633]{Elaine Winston}
\altaffiliation{SMOG mosaics and IRAC+MIPS catalog are available for download at: https://dataverse.harvard.edu/dataverse/SMOG\_CfA}
\affiliation{Center for Astrophysics  Harvard \& Smithsonian  \\
60 Garden St. \\
Cambridge MA 02138, USA}

\author{Joseph Hora}
\affiliation{Center for Astrophysics  Harvard \& Smithsonian \\
60 Garden St. \\
Cambridge MA 02138, USA}

\author{Robert Gutermuth}
\affiliation{Department of Astronomy \\
University of Massachusetts \\
Amherst, MA 01003, USA}

\author{Volker Tolls}
\affiliation{Center for Astrophysics Harvard \& Smithsonian \\
60 Garden St. \\
Cambridge MA 02138, USA}

\begin{abstract}

In this paper we undertake a study of the 21 square degree SMOG field, a {\it Spitzer} cryogenic mission Legacy program to map a region of the outer Milky Way towards 
the Perseus and Outer spiral arms with the IRAC and MIPS instruments. 
We identify 4648 YSOs across the field. Using the $DBSCAN$ method we identify 68 clusters or aggregations of YSOs in the region, having 8 or more members. 
We identify 1197 class Is, 2632 class IIs, 819 class IIIs, of which 45 are candidate transition disk objects, utilizing the MIPS 24 photometry.  
The ratio of YSOs identified as members of clusters was 2872/4648, or 62\%.    
The ratios of class I to class II YSOs in the clusters are broadly consistent with those found in the inner galactic and nearby Gould's Belt young star formation regions. 
The clustering properties indicate that the protostars may be more tightly bound to their natal sites than the class IIs, and the class IIIs are generally widely distributed. 
We further perform an analysis of the {\it WISE} data of the SMOG field to determine how the lower resolution and sensitivity of {\it WISE} affects the identification of YSOs as 
compared to Spitzer:  we identify 931 YSOs using combined {\it WISE} and 2MASS photometry, 931/4648 or 20\% of the total number identified with Spitzer.   
Performing the same clustering analysis finds 31 clusters which reliably trace the larger associations identified with the {\it Spitzer} data.    
Twelve of the clusters identified have previously measured distances from the {\it WISE} HII survey.  
SEDFitter modeling of these YSOs is reported, leading to an estimation of the IMF in the aggregate of these clusters which approximates that found in the inner galaxy, 
implying that the processes behind stellar mass distribution during star formation are not widely affected by the lower density and metallicity of the outer galaxy.  

\end{abstract}

\keywords{infrared: stars --- stars: pre-main sequence --- circumstellar matter}

\section{Introduction}

The study of star formation has been revolutionized with the launch of the {\it Spitzer} Space Telescope ({\it Spitzer}; \citet{werner}).  To date, most of the published 
studies have focused on star forming regions (SFRs) in the inner part of our own Galaxy: those regions located within $\approx$1~kpc of the Sun or towards the 
Galactic center/bar.  Regions such as Orion or the Gould Belt clusters cover sites of star 
formation from low mass clusters with hundreds of members to high mass regions with tens of thousands of young stellar objects (YSOs). They are within 1~kpc of the sun and can be studied 
in high resolution to substellar masses \citep{dunham, megeath}.    More distant SFRs located towards the galactic center, such as W51 and W43 (\citet{saral}, d$\sim$6~kpc) 
or RCW~38 (\citet{win11}, d$\sim$1.7~kpc), may be the precursors for young massive clusters (YMCs; \citet{bressert, ginsburg}).  {\it Spitzer} observations of these 
regions can detect YSOs down to $\sim$0.5-1$M_{\odot}$, and are more similar to what we can achieve towards the outer galaxy. 
Though the initial environmental conditions can vary greatly across these regions, properties such as the Initial Mass Function (IMF) appear to be universal across all formation size-scales.  
In calculating the star formation rates in external galaxies, it is generally assumed that this universality holds.   
However, when we look at extragalactic star formation, we are studying star formation not in just these inner galactic regions, but the entire integrated flux of the galaxy, which 
includes the equivalent of our own outer galaxy where we have not yet established the universality of the IMF slope.     

We expect that the efficiency with which a molecular cloud forms stars is dependent on its density, temperature, and chemical abundances \citep[e.g.][]{evans}.  
The outer galaxy is thought to be a distinctly different environment to that of the inner galaxy, with conditions less likely to efficiently form stars. 
The metallicity of the Milky Way has been shown to decline as a function of galacto-centric radius \citep[e.g.][]{rudolph}.  
A recent paper by \citet{huang} used red clump stars to show that [Fe/H] is decreasing in the midplane with radial distance from $R_{GC}$ of 7 to 11.5~kpc at $0.082 \pm 0.003$~dex/kpc, 
with $[Fe/H]\sim$~0 in the solar neighborhood.  There is less evidence for a decrease in the outer galaxy ($11.5 \le R_{GC} \le 14$kpc) with 
$[Fe/H]\sim$ $-0.2$ to $-0.4$ and $\Delta[Fe/H]$ of $-0.017 \pm 0.023$~dex/kpc. They ascribe this difference to 
different evolutionary histories in the inner and outer disk of the galaxy.

Average temperatures in molecular clouds are found to be lower \citep{mead}, and the cosmic ray flux is also thought to be lower \citep{bloemen}. 
Further, the volume density of molecular clouds in the outer galaxy is lower, and so interaction rates and incidence of spiral arm crossings will be lower over 
the star forming lifetime of the cloud.   
With the advent of all sky surveys we can now turn our attention to the outer Galaxy to identify new star forming clusters in these more remote regions to 
determine if the outer Galaxy is as amenable to the formation of stars as the inner Milky Way.  
For our purposes here, the term 'outer Galaxy' refers to clusters more than $\approx$2kpc from the Sun in directions away from the Galactic center from roughly { \bf $90^\circ < l < 270^\circ$. }

The {\it Spitzer} SMOG survey \citep{carey} was designed to provide deep coverage of a representative field in the outer Galaxy. 
It was observed during the cold mission, making it the largest field in the galactic plane of the outer galaxy with coverage across all IRAC and MIPS bands. 
We undertook an initial survey of the Outer Galaxy with this region, first, to make use of the longer wavelength coverage, and second, to then compare 
the YSOs identified with {\it Spitzer} and {\it WISE}.  Our future work will expand on this study by applying the techniques outlined in this paper to the outer Galaxy 
covered by the GLIMPSE 360 galactic plane survey (warm mission, IRAC bands 1 \& 2 only) and {\it WISE}.

\begin{figure*}
\epsscale{1.2}
\plotone{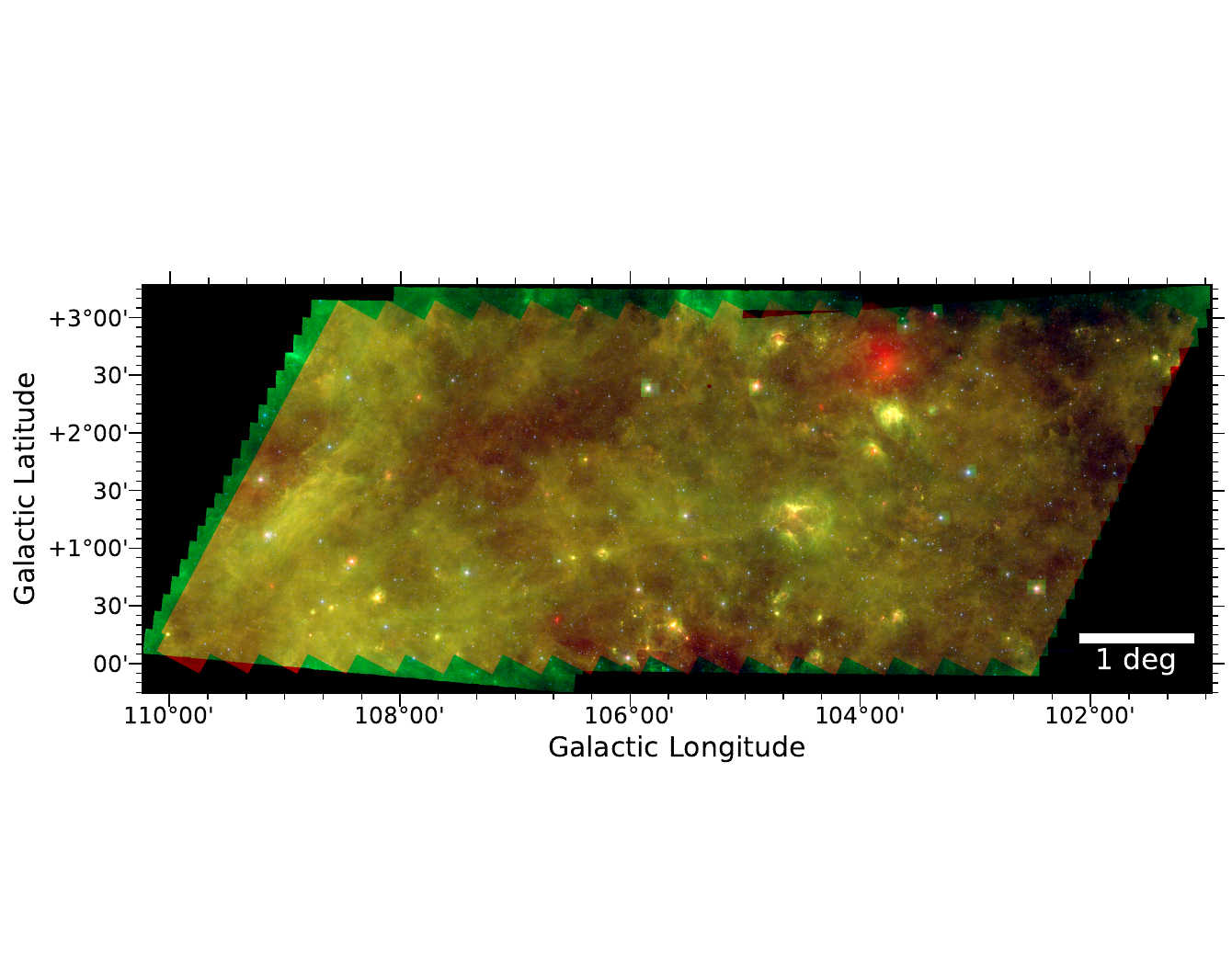}
\caption{ 
Spitzer IRAC three-color image of the full 21 square degree SMOG field, with blue: 3.6~${\mu}m$, green: 4.5~${\mu}m$, red: 8.0~${\mu}m$.
Dust emission dominates the field, with the larger brighter regions corresponding to areas of star formation activity.  
 }
\protect\label{fig_image}
\end{figure*}

With this work we identify young stellar objects (YSOs) across the field from their excess emission in the Infrared, locate new regions of star formation,  
determine the evolutionary class of the YSOs to analyze the protostellar ratio of the clusters, and make a preliminary assessment of the initial mass function 
(IMF) across clusters with known distances in the outer galaxy.   
Figure~\ref{fig_image} shows an IRAC three-color mosaiced image of the entire SMOG field, with the 8.0$\mu$m (red) highlighting a number of young 
stellar clusters, and the warm dust across the field, 4.5$\mu$m (green) tracing shocked hydrogen emission, and 3.6$\mu$m in blue.

In this paper we first discuss, in Sect.~\ref{obs}, the origins of the infrared catalogs.   
In Sect.~\ref{iding} we discuss the contamination removal process, the identification of the YSOs, and their evolutionary classification. 
We then discuss the spatial distribution of the young stars and the identification of stellar clusters in the outer galactic fields in Sect.~\ref{spatial}. 
We discuss the fits to the SEDs of YSOs in clusters with known distances in Sect.~\ref{sedfitter}.  We then compare the {\it WISE} catalog with the IRAC 
observations in Sect.~\ref{wise_compare}.   Finally, a brief summary is presented in Sect.~\ref{summ}.

\section{Observations and Data Reduction}\label{obs}

\subsection{SMOG Survey}

The {\it {\it Spitzer} Mapping of the Outer Galaxy} \citep[SMOG;][Prog. ID 50398]{carey}  survey was a {\it Spitzer} Cryogenic Cycle 5 program to map a 3$^\circ$ by 7$^\circ$ region of the outer Milky Way 
with IRAC \citep{fazio} and MIPS \citep{rie}.  The field covers 21 square degrees over  $102^\circ < l < 109^\circ$ and $0^\circ < b < 3^\circ$, 
covering the FWHM of the gas disk at R$\sim$20~kpc, and tracing the warp of the galactic plane in the second quadrant \citep{wouterloot}.   The field was selected to 
contain sections of two of the galaxy's spiral arms:  the Perseus arm and the Outer Spiral arm.  The Perseus arm lies at an estimated distance of d$\approx$3-5~kpc or 
galactocentric distance of R$\approx$9~kpc, while the Outer Spiral arm is located at d$\approx$7-10~kpc or R$\approx$12~kpc \citep{churchwell, xu}.

The SMOG data products, including mosaics and point source archive, were downloaded from the Infrared Science Archive (IRSA).  
The IRAC observations at 3.6, 4.5, 5.8, 8.0$\mu$m were obtained in High Dynamic Range mode, which obtained 0.6 and 12 second integrations 
for four dithered images at each mosaic position.  These data products were processed using the IRAC SSC S18.5.0 pipeline.   The data products consist of 
the SMOG Catalog, with 2,512,099 high reliability point sources, and the SMOG Archive, with 2,836,618 sources with less stringent selection criteria.  
The Archive has lower signal-to-noise thresholds and fluxes, and sources with separations as close as 0.5" are retained.  
In this study we utilize the larger Archive (referred to as {\it the SMOG catalog} going forward) to include as many faint YSOs as possible.  ]
The mid-IR photometry was supplemented by $J$, $H$ and $K$-band photometry from the 2MASS point source catalog \citep{skr}.   
The photometric catalogues were merged using a 1$\arcsec$ matching radius.  Documentation describing the SMOG survey reduction is available on the IRSA 
website\footnote{https://irsa.ipac.caltech.edu/data/SPITZER/GLIMPSE/doc/ smog\_dataprod\_v1.1.pdf}.
The data products also include the SMOG Image Atlas, consisting of 3.1$^\circ$ by 3.7$^\circ$ mosaics in each of the four IRAC bands, with a 1.2$\arcsec$ pixel size, which were 
merged using $Montage$ to create the image in Fig.~\ref{fig_image}.  

The MIPS mosaics (24, 70, 160$\mu$m) were taken at medium scan rate with full-width scan stepping.   
 Due to the high background and low angular resolution of the 70 and 160$\mu$m bands, point source extraction was only applied to the 24$\mu$m mosaic.   
The MIPS 24$\mu$m observations were reduced following \citet{gut08, gut09} and yielded a point source catalog containing 15298 objects.   
We merged the MIPS catalog with the SMOG Archive, requiring a 1$\arcsec$ matching radius, consistent with the positional accuracy of the datasets 
and to minimize false matches.   A match to an IRAC source was found for 14695 objects (96.1\%) of 
the MIPS sources.

\subsection{UKIDSS}

While the 2MASS survey covers the full sky, it only does so to a depth of 16.5~mag at J-band.  In order to improve our ability to detect fainter and more 
embedded YSOs and to add photometric coverage for the SEDFitter modeling, we have included the deeper {\it UKIRT Infrared Deep Sky Survey} (UKIDSS) 
near-IR J, H, and K$_s$ band photometry, with a depth of $\sim$19.8~mag at J-band and a FWHM of 0.9$\arcsec$. 
The UKIDSS project is defined in \citet{lawrence}. UKIDSS uses the UKIRT Wide Field Camera (WFCAM; \citet{casali}). The photometric system is described 
in \citet{hewett}, and the calibration is described in \citet{hodgkin}. The pipeline processing and science archive are described in \citet{hambly}.  
The UKIDSS survey does not cover the entire SMOG field, instead covering approximately half of the survey area from $l = 102^{\circ}$ to $l = 105^{\circ}$.  
The data were downloaded from the WFCAM Science Archive (WSA) as part of DR10.  
The catalog of sources covering the SMOG field contained 163,785 point sources.

\subsection{WISE}

The Wide-field Infrared Survey Explorer \citep[{\it WISE};][]{wright} provides mid-IR photometry  at 3.4, 4.6, 12, and 22$\mu$m.  
The AllWISE catalog is an all-sky survey combining the cryogenic {\it WISE} All Sky survey and the initial release of the NEOWISE post-cryogenic survey \citep{mainzer}.  
The catalog is available via the NASA/IPAC Infrared Science Archive (IRSA) archive. 
A selection covering the entire SMOG field was made, resulting in a regional catalog of 581,705 point sources.   
The {\it WISE} satellite has considerably lower spatial resolution when compared to Spitzer, with a highest resolution of $\sim$6.1$\arcsec$ at 3.5$\mu$m.   
However, the astrometric accuracy and matching to the 2MASS catalog are to within 1$\arcsec$, and so we performed catalog matching to the SMOG catalog at 1$\arcsec$ also.

\section{YSO Identification \& Classification}\label{iding}

Young stellar objects are frequently identified by their excess emission at IR wavelengths. This emission arises from reprocessed 
stellar radiation in the dusty material of their natal envelopes or circumstellar disks. The infrared identification of YSOs is 
carried out by identifying sources that possess colors indicative of IR excess and distinguishing them from reddened and/or 
cool stars \citep{winston, all, gut1}.  

A full description of the criteria for contaminant removal and source identification for each dataset is given in the Appendices.  
The following subsections outline the YSO selection methods for each dataset, the combined YSO catalog, and the evolutionary classification of 
the YSOs based on their excess IR emission.

\subsection{IRAC \& 2MASS}\label{irysos}  

\subsubsection{Contamination}\label{} 

The SMOG field points towards the outer galaxy, where the background extinction from the Galactic bar is negligible. It is expected that 
many of the point sources detected with excess emission will, in fact, be active galactic nucleii (AGN) or star-forming galaxies (PAH galaxies).   Knots of emission in the 
structure of molecular clouds may also be mistaken for YSOs.  A further source of contamination in the YSO sample comes from 
contamination of the apertures by polycyclic aromatic hydrocarbon (PAH) emission.  These sources were identified in the SMOG catalog 
 and removed before selection of YSOs was carried out. 
 
\begin{figure*}
\plotone{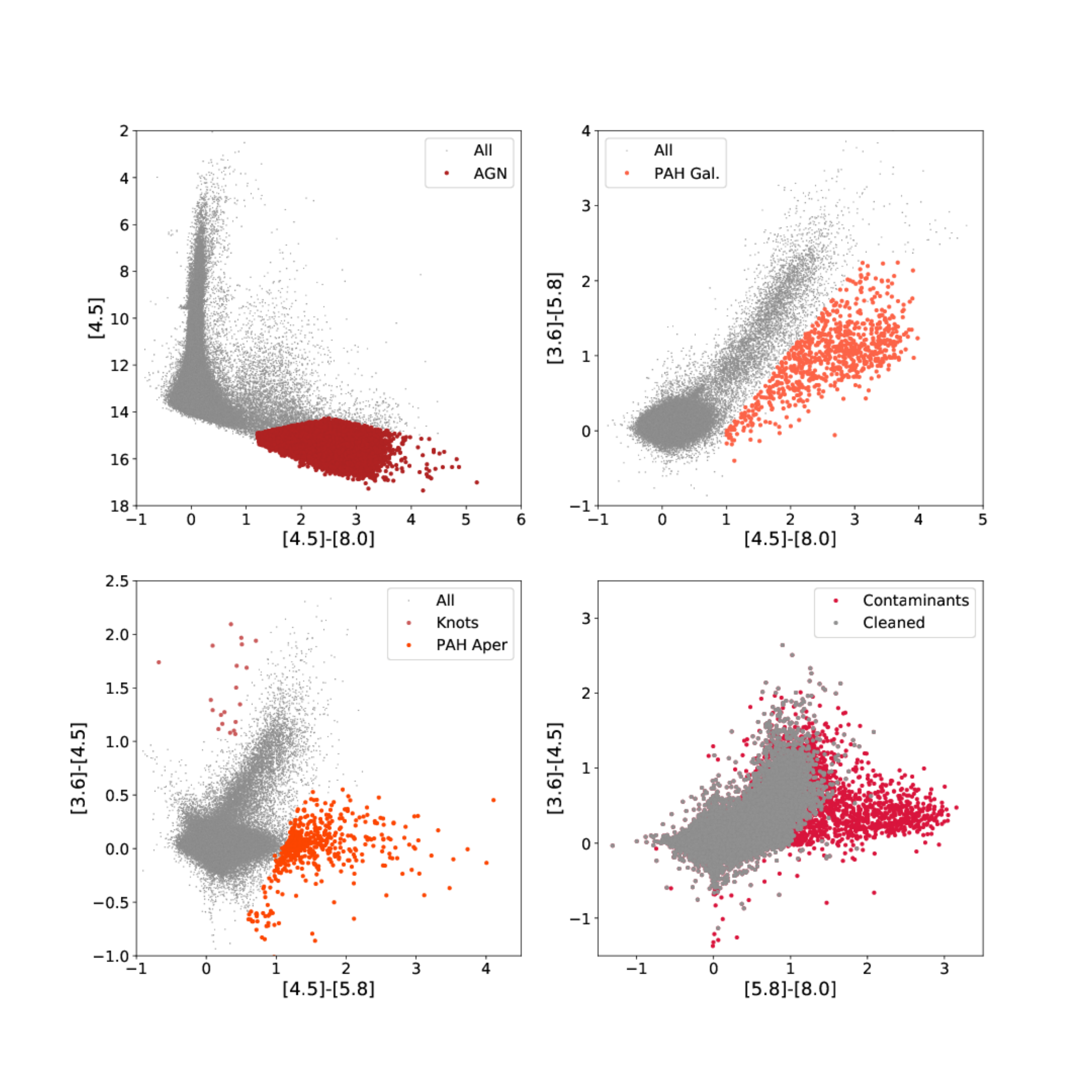}
\caption{ 
Selection of color-magnitude and color-color diagrams of the IRAC contaminants selection,  overlaid on all sources in the SMOG field. 
In each diagram the red/orange points indicate the selected contaminants.   
{\it Upper Left:} IRAC  [4.5] v [4.5 - 8.0], identification of AGN contaminants.   
{\it Upper Right:}  IRAC [3.6 - 5.8] v [4.5 - 8.0], identification of PAH galaxies.  
{\it Lower Left:}     IRAC [3.6 - 4.5] v [4.5 - 5.8], identifiation of knots and PAH contamination of the aperture.
{\it Lower Right:} IRAC [3.6 - 4.5] v [5.8 - 8.0], four band IRAC diagram showing the full catalog and the cleaned catalog after removal of contaminants.    
}
\protect\label{fig_contam}
\end{figure*}

Figure~\ref{fig_contam} shows the contaminant removal diagrams. In the first three plots the full catalog is displayed in gray, while the various contaminants 
are highlighted in red.  The final plot shows the IRAC color-color diagram (CCD) with the cleaned catalog in gray, overlaid on the full catalog in red.  
A total of 10,300 sources were removed from the catalog as contaminants. This cleaned catalog was then used to select the YSOs across the SMOG field.

\subsubsection{YSO Selection}\label{}

Young stellar objects were selected using a combination of CCDs with IRAC and 2MASS+IRAC colors, as described in Appendix~\ref{irysos}.
Photometric uncertainties of $< 0.2$~mags and saturation magnitude limits were required in all the bands used for a {\it particular} color-color diagram to select YSOs.  
Sources with colors and magntiudes consistent with YSOs were classified as class 0 / class I: 
for (deeply) embedded protostars,  class II: disk-bearing pre-main sequence objects, and class III: objects with weak/anaemic disks.  
 
\begin{figure*}
\plotone{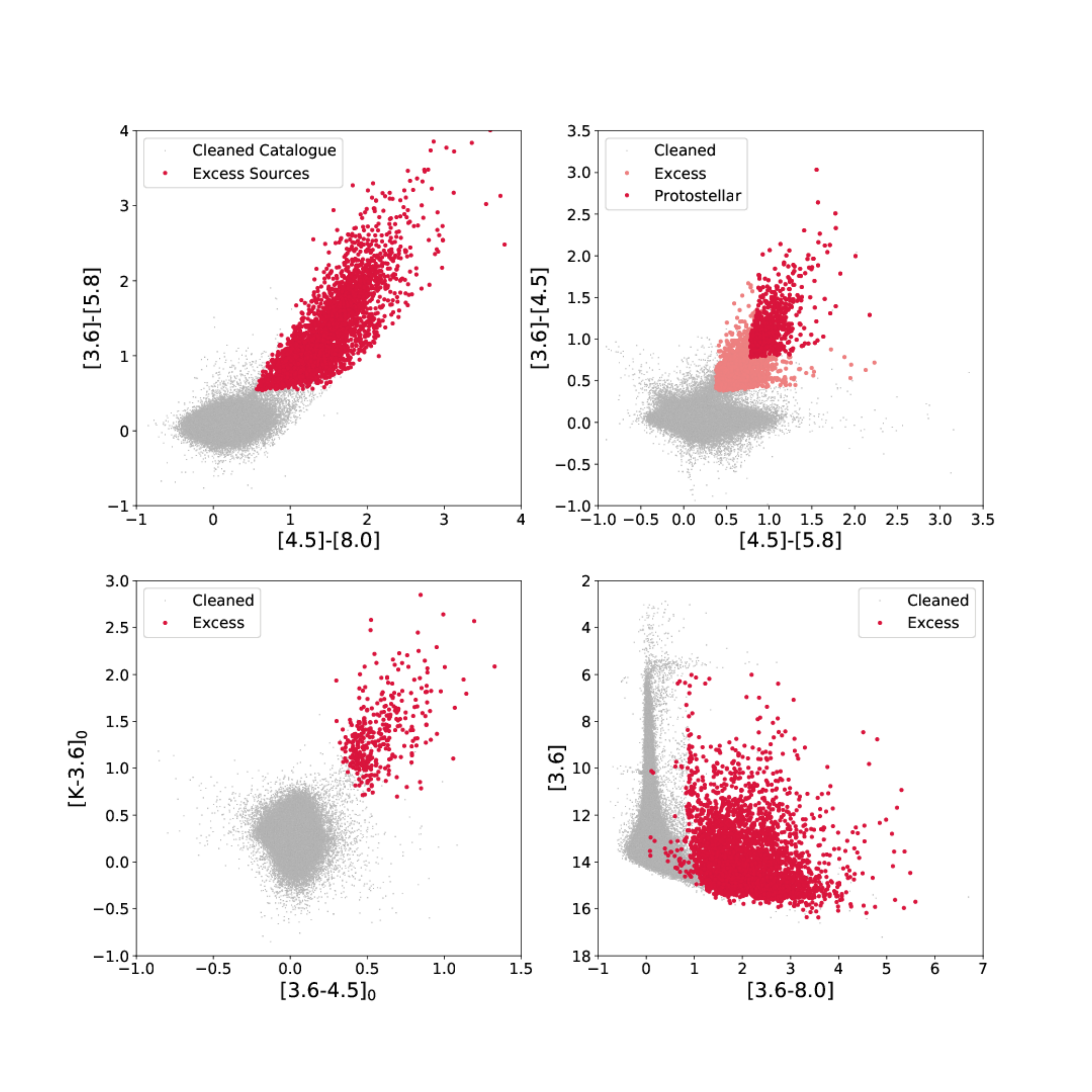}
\caption{ 
Selection of color-magnitude and color-color diagrams of the near- and mid-IR 2MASS and IRAC YSO selection (red circles),   
overlaid on all sources in the SMOG field (gray points).
{\it Upper Left:} Mid-IR four-band IRAC [3.6-5.8] v [4.5 - 8.0]: for sources with photometry in all four IRAC bands.    
{\it Upper Right:}  Mid-IR three-band IRAC [3.6-4.5] v [4.5 - 5.8]: for sources lacking photometry in the longest 8.0$\mu$m band.  
{\it Lower Left:} Near-IR 2MASS and IRAC [K - 3.6] v [3.6 - 4.5]: for sources lacking photometry in the two longest IRAC bands.      
{\it Lower Right:} IRAC [3.6] v [3.6 - 8.0]: illustrative diagram of distribution of identified YSOs in color-magnitude space.  
}
\protect\label{fig_irac}
\end{figure*}

There were 2,826,318 sources in the cleaned SMOG catalog, of which 3835 were identified as YSOs using the combined 2MASS and IRAC photometry.  
Figure~\ref{fig_irac} shows three of the source selection color-color diagrams used in the identification of YSOs, while the fourth shows a color-magnitude diagram 
of the cleaned catalog and the identified YSOs.  
Some contaminants that are not accounted for here include foreground and background stars, such as asymptotic giant branch (AGB) stars and highly/unusually reddened field stars,  
that may be confused for class III objects.  
Such objects are expected to be scattered randomly over the field, and some of the brighter YSO candidates may be foreground AGB stars. 
A more detailed discussion is given in Sec.~\ref{sec_sls} on large-scale clustering.

\subsection{MIPS}\label{mipsdata}

\begin{figure*}
\plotone{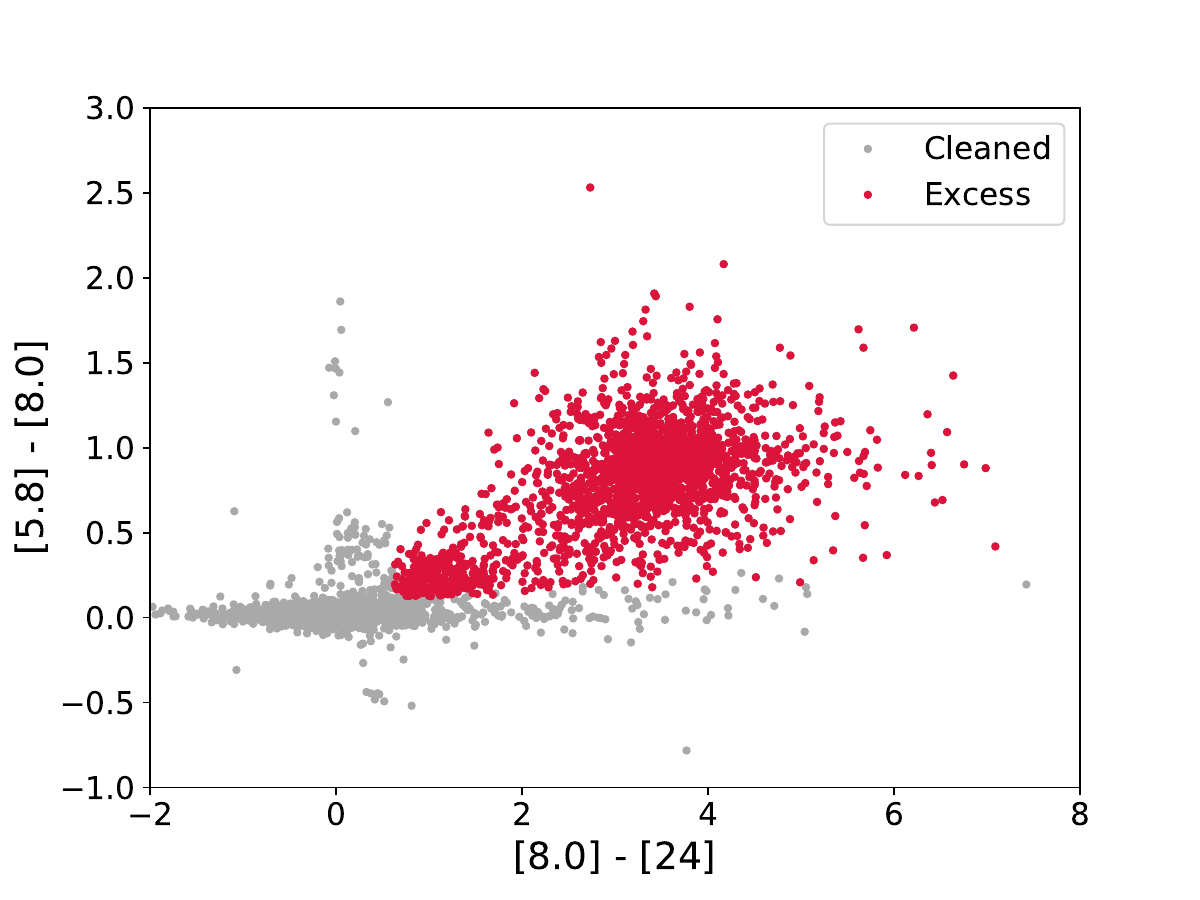}
\caption{ MIPS 24$\mu$m color-color digram showing the identification of YSOs using a combination of the longest two IRAC bands and MIPS band 1.  
This color combination is useful in selecting deeply embedded objects that may lack photometry at shorter wavelengths.  It is also useful in the 
preliminary selection of candidate transition disk objects.   }
\protect\label{fig_mips}
\end{figure*}

The MIPS 24$\mu$m mosaic yielded 15298 sources.  Of the 14695 matches to an IRAC counterpart, 14226 (96.8\%) had matches once the contaminant removal had been performed on the SMOG catalog.  
The remaining 603 unmatched sources, or 3.94\% of all the 24$\mu$m detections, are likely to be either background extragalactic sources, or perhaps deeply embedded protostellar objects.  
There were no sources with only 8$\mu$m and 24$\mu$m detections, and only 17 detected only at 5.8, 8.0, and 24$\mu$m.  The YSO selection criteria were thus based on two complimentary methods, one including only 
the three longest bandpasses (c.f. Fig.\ref{fig_mips}), and the second including the shorter IRAC bands and MIPS 24$\mu$m, c.f. Appendix~\ref{mipsysos}.  
These methods identified 2109 and 1819 sources, respectively, for a combined total of 2140 YSOs identified with 24$\mu$m excess.  

An attempt was made to identify any embedded protostellar objects from the list of 24$\mu$m only sources.  This was done by examining their spatial distribution across the field 
in relation to other identified YSOs.  The objects were not found to be spatially coincident with the other YSOs, instead being randomly distributed, and so no further attempt is made to classify them here, though the possibility remains that some are deeply embedded protostars.

\subsection{IRAC \& UKIDSS}\label{}

\begin{figure*}
\epsscale{0.8}
\plotone{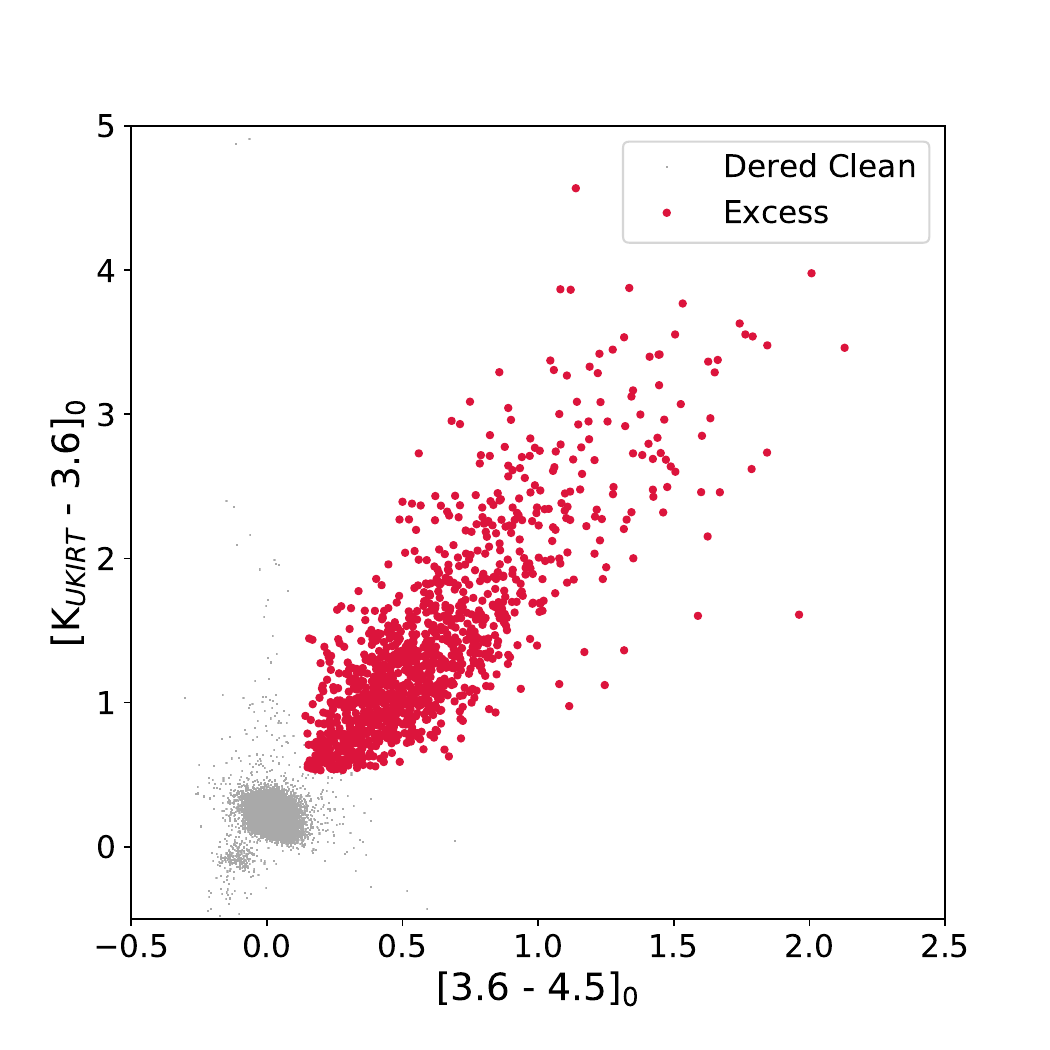}
\caption{ Color-color diagram including the UKIRT near-IR photometry in place of 2MASS in the [K - 3.6] v [3.6 - 4.5] yso selection.  
The UKIRT survey covered $\sim$50\% of the SMOG field to a greater depth than the all-sky 2MASS survey.   The photometry was dereddened 
prior to source selection to remove line of sight interstellar dust extinction.   }
\protect\label{fig_ukirt}
\end{figure*}

The UKIDSS near-IR photometry were merged with the cleaned SMOG catalog using a 1$\arcsec$ matching radius. 
The rms positional uncertainty for IRAC is $<0.3\arcsec$ after pointing refinement, so matching at 1$\arcsec$ allows 
 for a 3-$\sigma$ positional offset while incorporating positional uncertainties in the UKIDSS photometry.   
 If more than one object falls within the 1$\arcsec$ radius, the closest is taken to be the match.
Of the 163,785 UKIDSS sources, 163,485 were matched to a mid-IR detection.  No further contaminant removal was applied to the SMOG data.  The UKIDSS photometry is saturated for $K  \gtrapprox 10.5$, and so it was necessary to apply photometric 
cuts to the matched sources to avoid selecting saturated sources.  We expect all brighter sources to have been previously detected with 2MASS.   
Figure~\ref{fig_ukirt} shows one of the selection criteria used to identify YSOs with the combined IRAC and UKIDSS dataset.  There were 1182 and 979 sources identified, 
respectively, with the two criteria, c.f. Appendix~\ref{ukysos}.  A combined total of 1352 YSOs were selected.  As previously mentioned, the UKIDSS coverage consisted of roughly half of the SMOG field.  
For this reason our survey is deeper over one half of the field, and YSO completeness is higher for young objects with fainter near-IR emission in this region.

\subsection{Combined YSO Catalogue}\label{irysos}

The three sets of YSOs selections, 2MASS+IRAC (3835 sources), IRAC+MIPS (2140 sources), and UKIRT+IRAC (1352 sources), were combined using source identification tagging.  
All three catalogs required a detection in the SMOG catalog, which uniquely identifies each source.  These were combined and overlaps removed.  
The full list of YSOs contained 4648 objects across the SMOG field.  

Table~\ref{tab:IRphot} lists the column identifiers of the photometry table for the full list of identified YSOs.  The full data table for all 4648 YSOs is available online in electronic format.  
The portion of the SIMBAD database covering the SMOG field was downloaded and the sources compared to the locations of the 4648 identified YSOs.  
Table~\ref{tab:Simbad} provides a sample listing of the YSOs with matches within 1$\arcsec$ of a SIMBAD source, the source identification, the object type, and the spectral type. 
The full table is available online in electronic format, and includes further selected information pertaining to the SIMBAD objects.

\subsubsection{Evolutionary Classification}

Young stars evolve through a number of broad stages from the embedded core phase, through the protostellar phase where stellar accretion is still dominant, 
to the circumstellar disk phase where the envelope has dissipated and processing of disk material is ongoing, to the weak disk regime where the disk has 
dissipated and any planets will have formed.  

A broad classification may be carried out by measuring the slope, $\alpha$, of the spectral energy distribution (SED) across the mid-IR bandpasses \citep{lad84}.  
Two values of the slope were calculated by taking the least squares polynomial fit to the data from 3.6-8.0$\mu$m (IRAC) and from 3.6-24.0$\mu$m (IRAC+MIPS1).   
Protostellar objects (Class 0 and I) have a rising slope, $\alpha > 0$;   class II sources are characterized by decreasing slopes between$-1.6 < \alpha < -0.3$, 
while class III sources lack optically thick emission from a disk and possess decreasing slopes $\alpha < -1.6$, consistent with a weak emission above a stellar photosphere.   
We note a distinction in nomenclature between class III sources as defined here and X-ray detected class III YSOs \citep{win10, win11}.
Here class IIIs show weak IR emission consistent with a thin disk or weak line T-Tauri star, while X-ray detected diskless class III sources do not show any excess IR emission 
in the IRAC and MIPS bands and would have $\alpha \approx -2$ and be classified as field stars by our method.  

The $\alpha_{IRAC}$ classification was chosen as it can be applied to the entire combined YSO list simultaneously, regardless of how the YSO was selected.  
The YSOs were classified as follows:  1197 class 0/I,  2632 class II sources, while 819 were identified as anaemic disk class III stars.   Of the 
819 class IIIs,  45 objects were further identified as candidate transition disk objects based on their possessing rising slopes between 8$\mu$m and 24$\mu$m.

\subsubsection{Incompleteness of the YSO Population}

An important topic to address in a survey aimed at identifying a specific population is the subject of incompleteness.  
In this case, we can address the incompleteness both in terms of total YSO membership and of evolutionary class.  

The primary difficulty in discussing incompleteness in the total cluster populations, in the context of this study, is that 
the field contains clusters spanning a wide range of distances, from 1-10kpc, through two of the Galaxy's spiral arms. 
Thus we cannot make a meaningful estimate of the minimum mass of YSO detected without referring to a specific cluster;  
we can only discuss the incompleteness in terms of the photometric cuts applied as selection criteria. As we do not have 
distances to all of the identified clusters, and the available distances are not confirmed, we make no further assessment 
of the completeness of the catalog of identified YSOs in this study.

In discussing the incompleteness by evolutionary class we can say: we are more sensitive to Class II sources, which 
are less embedded and are dominated by photospheric emission, and therefore tend to be brighter in the shorter IRAC bands 
than the protostellar Class I sources. The more deeply embedded Class Is may be detected by MIPS, but here the lower 
resolution and source crowding will reduce the selection of Class Is in dense and/or more distant clusters.  
To discuss the incompleteness of the Class III objects, the distinction between truly diskless Class III YSOs and those 
with weak/anaemic disks must be taken into consideration.  Those with weak disks, as identified here, will have a similar 
completeness to the Class IIs.  Class IIIs without disks will be completely missed as they cannot be distinguished from 
field stars at IRAC wavelengths.  Spectroscopic or X-ray observations  would better allow us to identify the 
full Class III populations of these regions.

\section{Cluster Identification}\label{spatial}

\subsection{Large Scale}\label{sec_sls}

\begin{figure*}
\epsscale{1.2}
\plotone{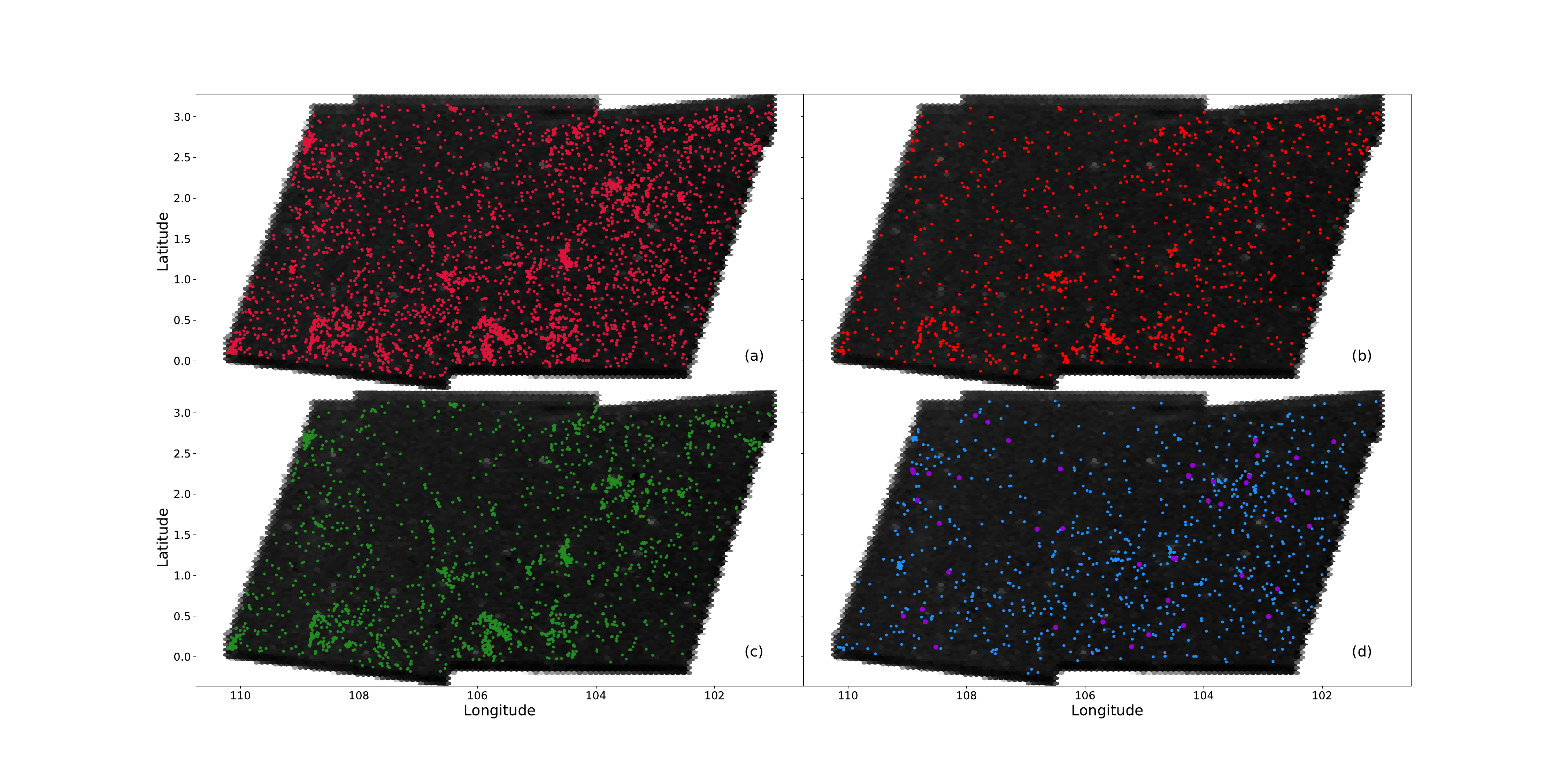}
\caption{ Spatial distribution of  (a) all identified YSOs in the SMOG field from the combined 2MASS-IRAC, UKIRT-IRAC, IRAC, and MIPS photometric 
selection criteria, (b) showing only the protostellar objects, classified as class I, (c)  showing only the pre-main sequence objects, classified as class II, 
(d) showing only the more evolved objects, classified as class III or candidate transition disks.
\label{fig_sda}}
\end{figure*}

The SMOG survey covers a 21 square degree field towards two of the outer spiral arms of the Milky Way galaxy.  A cursory visual examination of the spatial distribution 
of the YSOs shows evidence of clustering/clumping, as can be seen in Figure~\ref{fig_sda}.  The spatial distribution of all identified YSOs is shown in Fig.~\ref{fig_sda}(a), 
subsequent three plots in Fig.~\ref{fig_sda} show the spatial distributions of the class I (panel b), class II (c), and class III (d) with candidate transition disks, respectively.  
The class II sources most clearly show the structures of their underlying cluster distributions.  The class I YSOs show a similar structure, though they are less numerous and 
somewhat more tightly associated than the class IIs.  The class IIIs do not show as strong a trend towards location in the clustered regions as the younger objects, though 
in one exception there is a knot entirely consisting of class III sources.  

In order to identify over-densities in the spatial distribution of the identified YSOs, and thus pick out 'clusters' or 'regions' of star formation, we elected to use a method 
called "Density-based spatial clustering of applications with noise" ($DBSCAN$) as described by \citet{ester}.  This is a density-based clustering algorithm, where points 
with many nearby neighbors are selected to form a group and those with only more distant neighbors are flagged as outliers.  The algorithm used was the 
Python $SKLEARN$ package's implementation of $DBSCAN$.  The method requires only two input parameters: 
$\epsilon$, the scaling size for clustering, and $MinPts$, the minimum number of points required to define a dense region.  
A detailed description of $DBSCAN$ and its utility in finding clusters of YSOs in SFRs is presented by \citet{joncour} in their paper on Taurus.  
Here, the value of $\epsilon$ was chosen by plotting the cumulative distribution of nearest neighbor distances for all the YSOs, and determining the turn-off point of the distribution.  
A value of $\epsilon_{tp} = 0.11\deg$ was selected from this plot.  The minimum number density of group members was set at $MinPts = 8$, to minimize the number of false clusters detected.   

\begin{figure*}
\epsscale{1.3}
\plotone{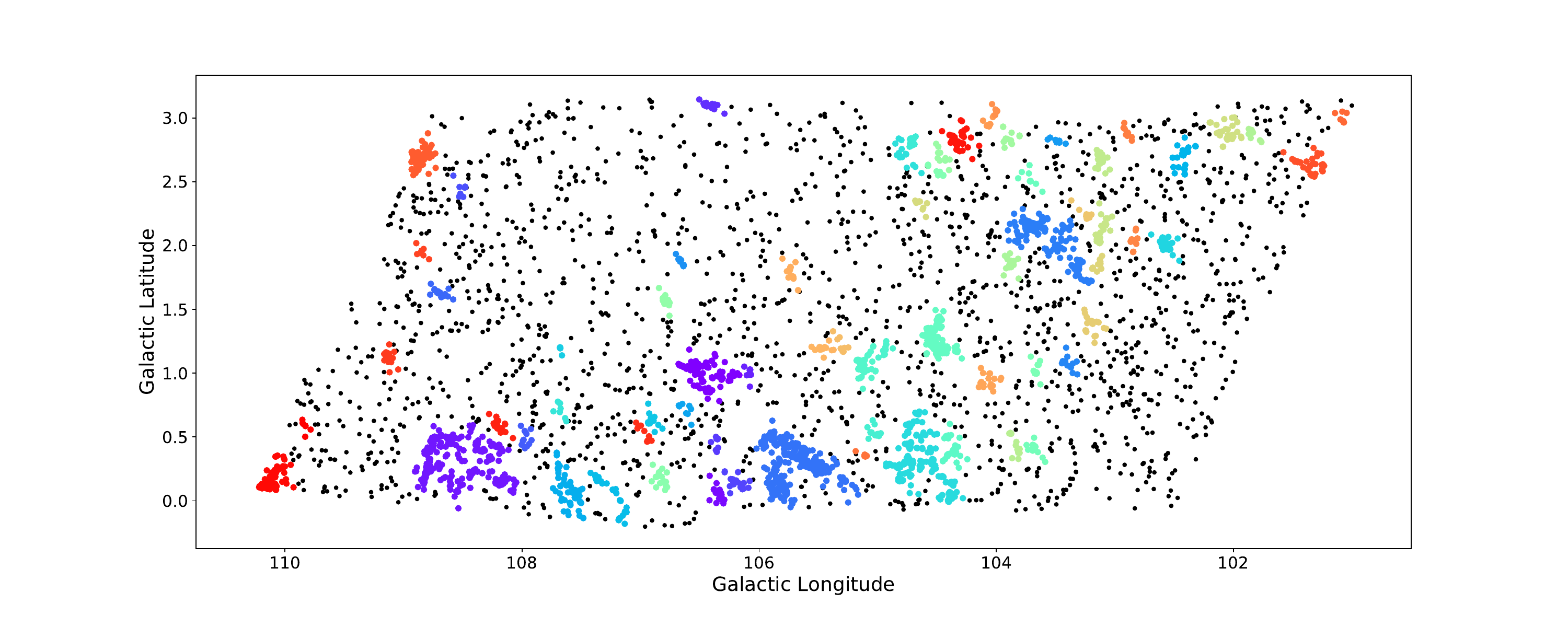}
\caption{ Spatial distribution of the identified YSOs showing the clusters identified by the DBScan method.  The sources in each cluster are color-coded.  
The black dots represent those YSOs not identified as belonging to a cluster.   }
\protect\label{fig_dbscan}
\end{figure*}

Using this method, 68 clusters or groupings were identified.  Figure~\ref{fig_dbscan} plots the spatial distribution of all identified YSOs in black, with the colored dots 
representing the identified clusters.   The smallest identified cluster contained 8 members, and the largest cluster contained 429 members, with a median size of 17 members.  
Eight clusters had 80 or more members:  Clusters \#0,   \#2,   \#10,  \#11,  \#22,  \#30,  \#59, and \#66, with 155, 318, 428, 198, 221, 174, 80, and 104 members, respectively.   
A point to note is that the addition of the UKIDSS near-IR data over one half of the field leads to more YSOs detected in that area, which in turn leads to 
an increase in the number of small-sized clusters detected on that half of the field that would not be detected otherwise.  
Of the 68 clusters identified,  9 had 8 members, 45 had 25 members or less, and 60 had 80 members or less.

To determine the statistical likelihood that the identified groupings are bona fide stellar clusters, we repeated the $DBSCAN$ analysis for 1000 iterations with distributions of randomly 
generated  points covering the same area as the SMOG field.  The size of the distribution was tied to the value of $MinPts$ so that it equalled the number of YSOs in the field and in 
clusters at and below that value.  We found that setting the $MinPts$ number density to 8, returned one false cluster 0.1\% of the time. 
For a $MinPts$ number density of 6-7, fake clusters were detected $\approx$20\% of the time, rising to $\approx$50\% when that value drops to 5.  Therefore we assume that all 
identified clusters of 8 members or more are real, and hence all 68 clusters identified in the SMOG field are likely to be bona fide stellar associations.  
The choice of value for  $\epsilon_{tp}$ was tested with a range in values around $0.9 < \epsilon < 0.13 \deg$  to assess 
the reliability of the cluster identification method. The number of clusters identified was [74, 70, 70, 68, 71, 70, 65] for 
values of $\epsilon$ of [0.09, 0.1, 0.105, 0.11, 0.115, 0.12, 0.13], respectively. In all cases, the central cores of the clusters 
identified remain the same, while some YSOs are added or removed.  As $\epsilon$ decreases, the larger clusters fragment into multiple 
smaller clusters that are identified as subclusters, c.f. Sec.~\ref{sec_ssc}.  As $\epsilon$ increases, regions merge and more 
diffuse regions are identified as clusters, eventually leading to the merger of all clusters in the field as $\epsilon >> \epsilon_{tp}$.   
From this analysis, we have selected  $\epsilon_{tp} = 0.11\deg$  as the turn-off point in the cumulative distribution.

The total number of YSOs identified as belonging to a cluster was 2888 out of 4648, or 62\%, or conversely, that 1776 out of 4648, or 38\% of the YSOs were outliers. 
It is possible that some of these outliers are associated with clusters/associations located nearby to them but fainter YSOs connecting them are not detectable with 
this dataset.   They may also be contaminants; foreground or background galactic field dwarfs or AGB stars with either a high extinction or a small amount of dusty 
material surrounding them that leads to an excess of flux in the mid-IR.    Of the 1776 outliers,  512 were class I, 748 were class II, and 516 were class III objects.    

Two approaches were taken to estimating the AGB contaminant population:  Using the {\it Besancon} galactic population synthesis models 
\citep{robin03, robin14}, we obtained the modeled population in one square degree towards $l = 105$ and $b = 1$, and found 2 AGBs, which when scaled to 21 sq. deg yields 
a population of roughly 40 AGB stars along the line of sight of the SMOG field.
 A second method of estimating the AGB population was described in \citet{robitaille2008}, by modeling the Galactic population of AGB stars based on that found in the LMC \citep{srinivasan}.  
 By applying a color cut of $[8.0 - 24] < 2.5$ to the YSOs with a 24$\mu$m detection, we estimate a population of approximately 400 AGB stars in the SMOG field. 
 We can consider these two values as upper and lower limits to the AGB contamination in the field.
 Thus, it is possible that a fraction (23\%) of the non-clustered candidates identified are not YSOs, but are in fact Galactic AGB stars.

\begin{figure*}
\epsscale{1.2}
\plotone{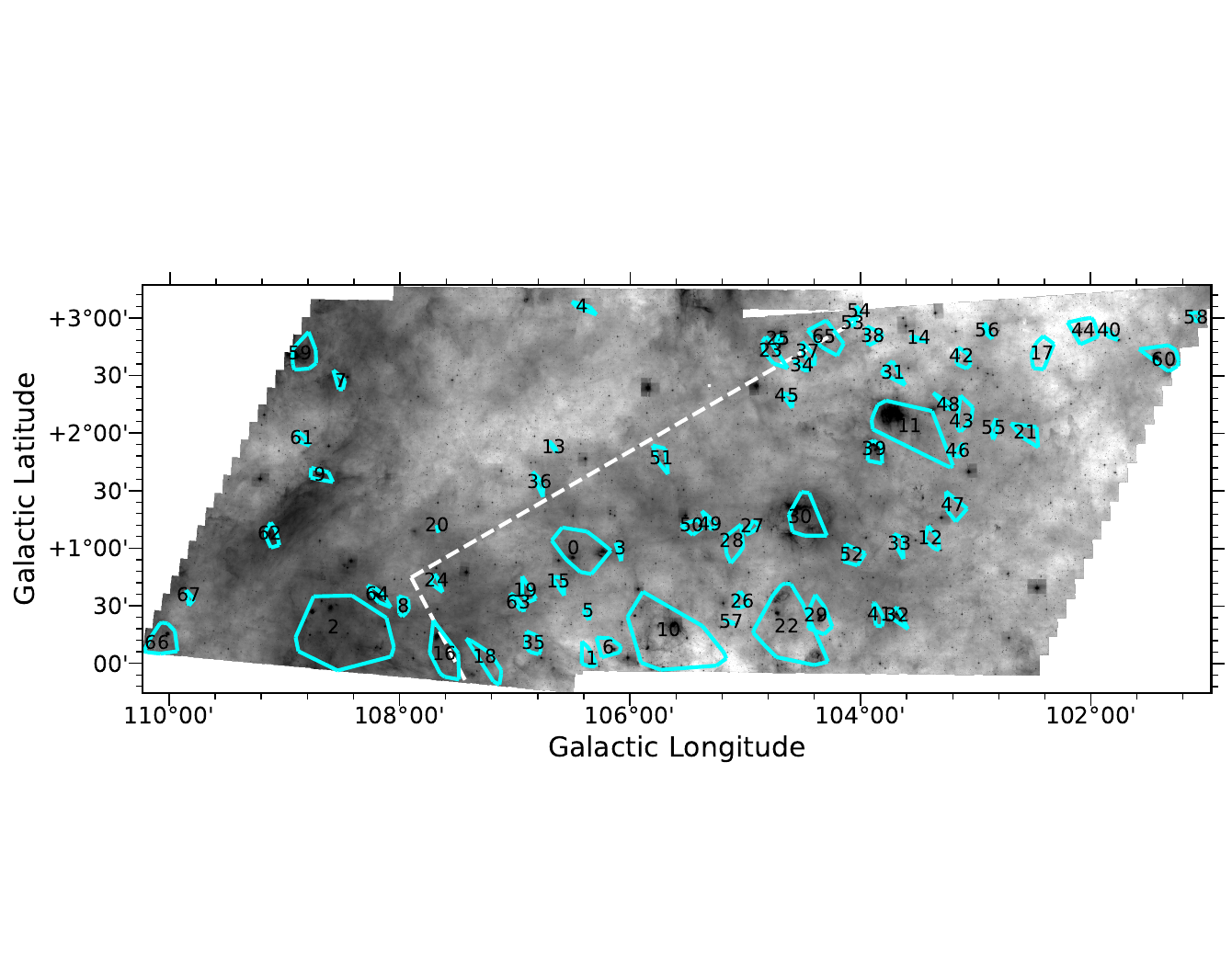}
\caption{ Grayscale showing the IRAC 8.0$\mu$m 
mosaic of the SMOG field.  The cyan contours overlaid on the image represent the calculated 
convex hulls for each cluster, providing a rough spatial outline of each cluster, annotated with the cluster number. 
 The area to the right (lower longitudes) of the dashed white line is covered by the UKIDSS photometry.}  
\protect\label{fig_greycluster}
\end{figure*}

\begin{figure*}
\gridline{\fig{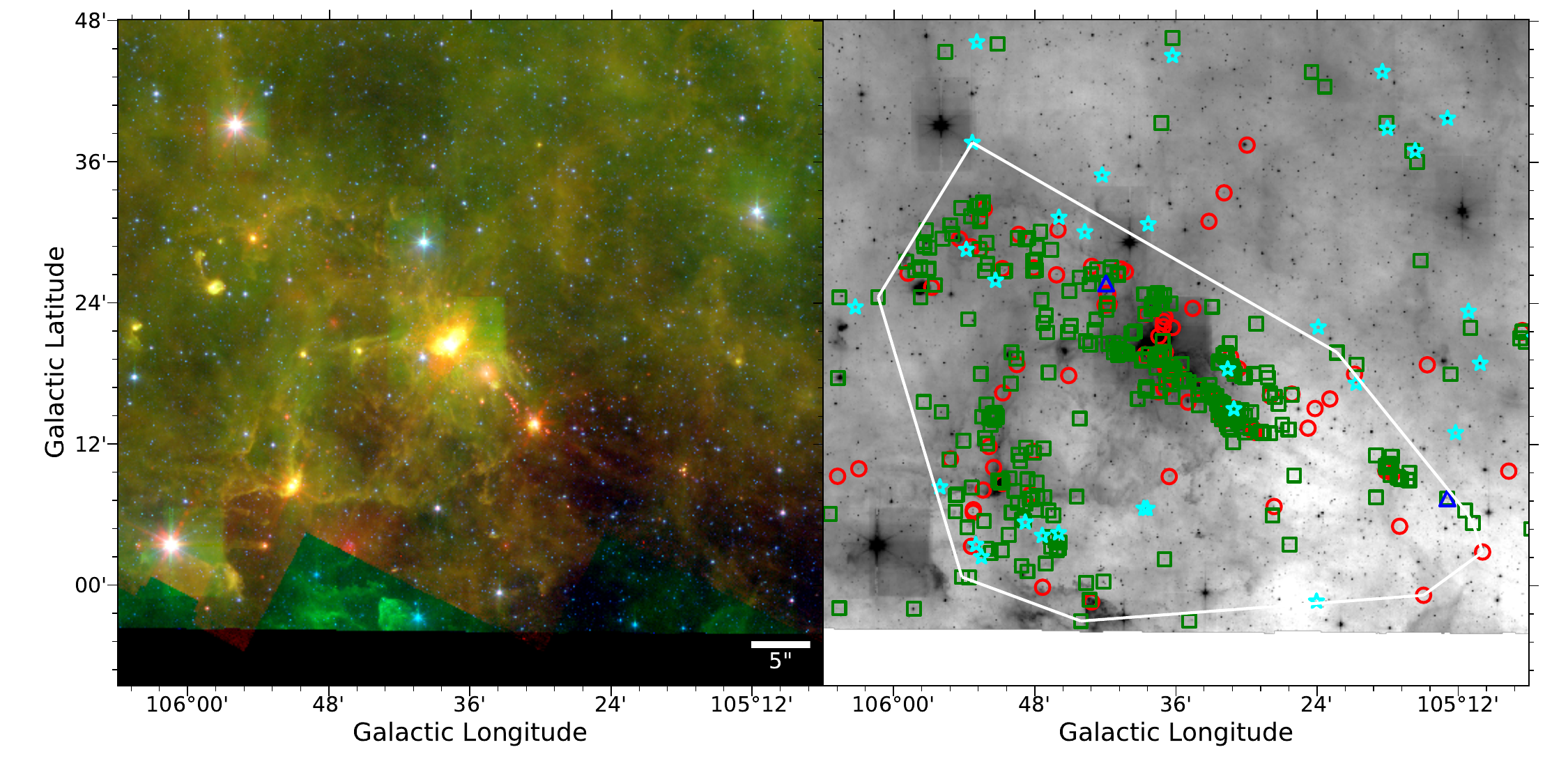}{0.75\textwidth}{}  }
\gridline{\fig{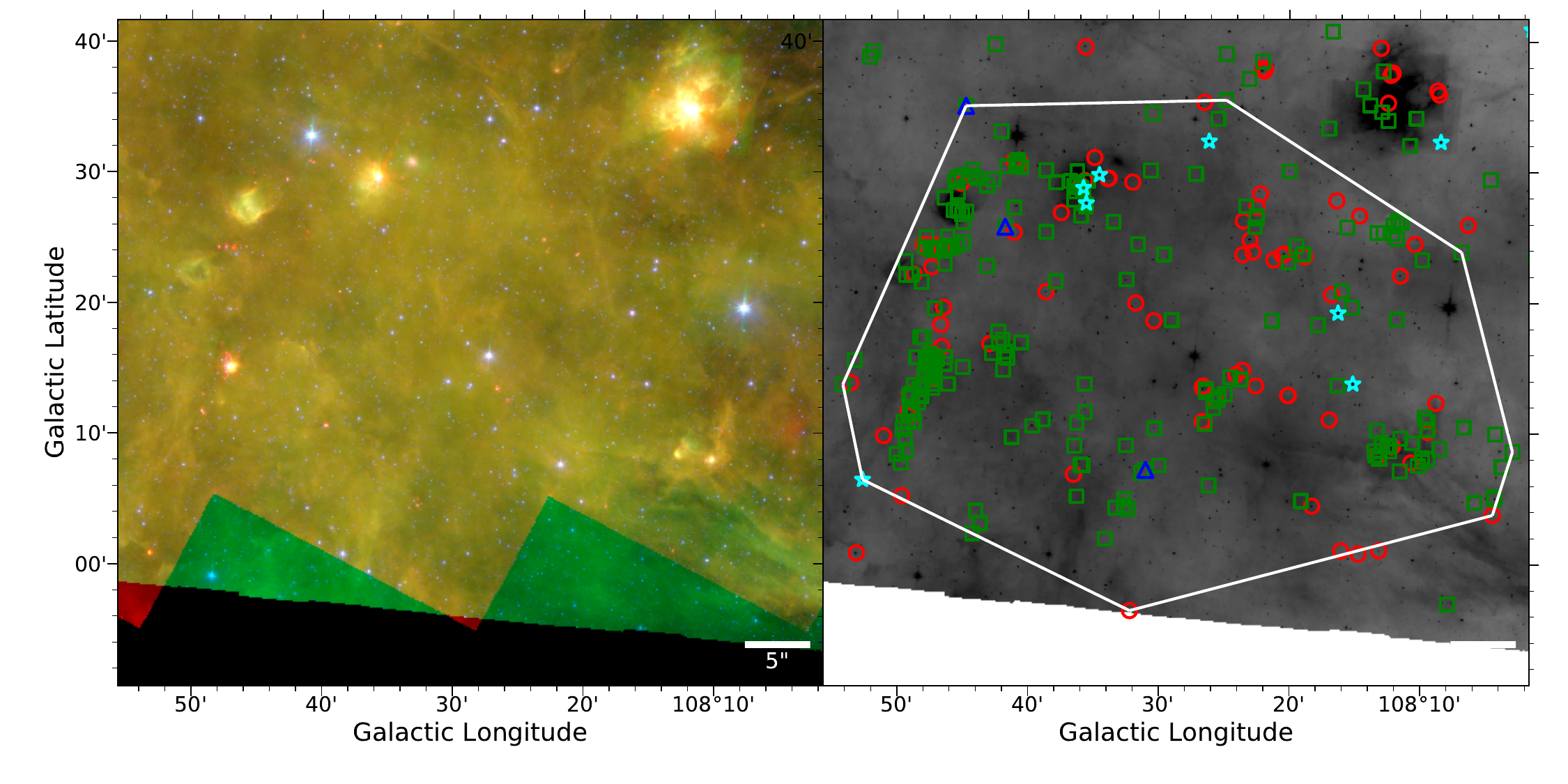}{0.75\textwidth}{}  }
\gridline{\fig{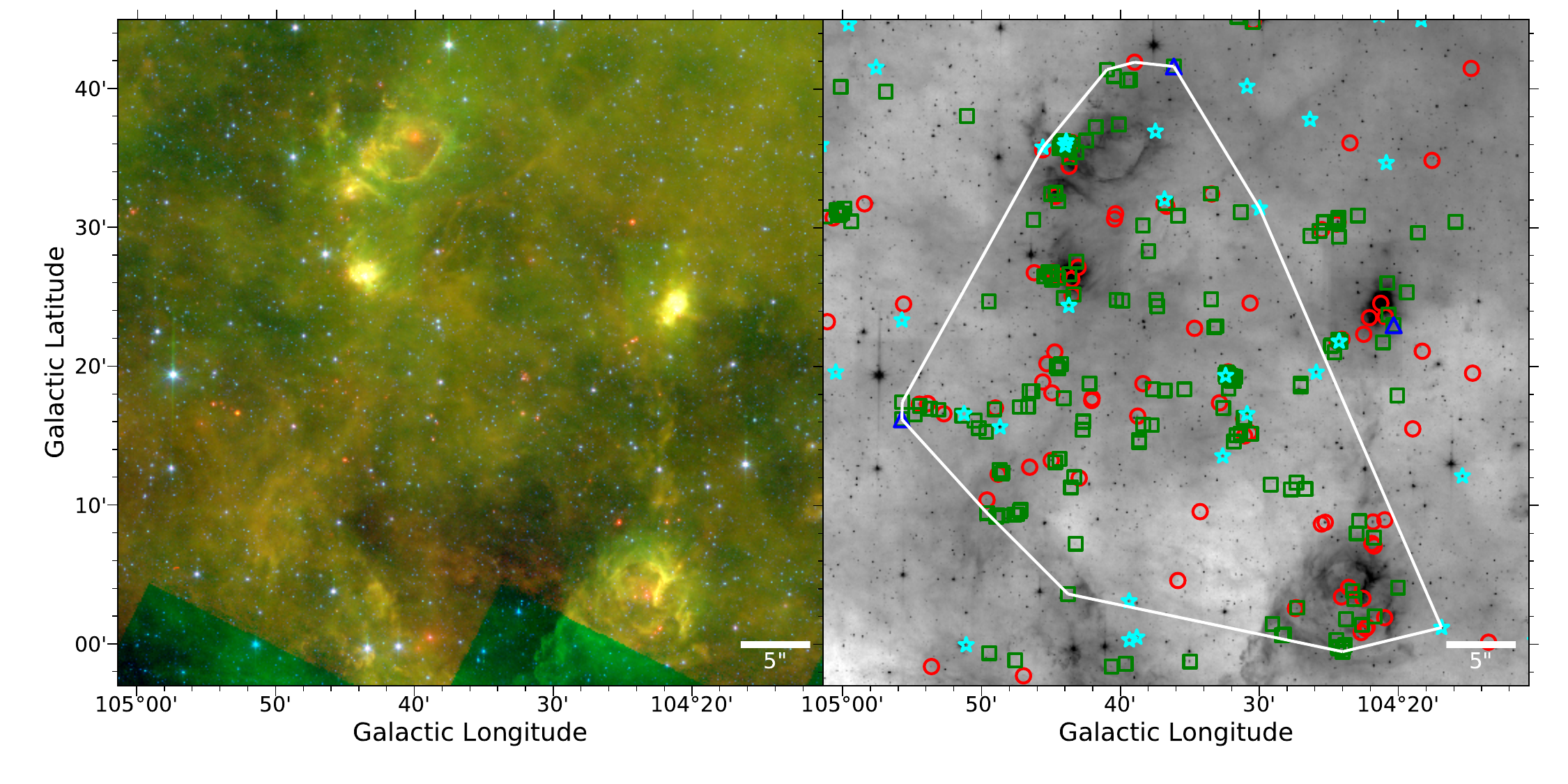}{0.75\textwidth}{}  }
\caption{ The first three of the eight largest clusters identified in the SMOG field by cluster membership from largest to smallest: \#10, \#2, \#22 
from top to bottom, in 3-color (r: 8.0, g:4.6, b:3.6$\mu$m) and IRAC 8.0$\mu$m grayscale.  
The contours overlaid on the grayscale represent the calculated convex hulls for each cluster, providing a rough spatial outline of each cluster.  
The symbols represent the YSOs; class I (red circle), class II (green square), class III (cyan star), transition disk (blue triangle).    }
\protect\label{fig_grey8_1}
\end{figure*}

\begin{figure*}
\gridline{\fig{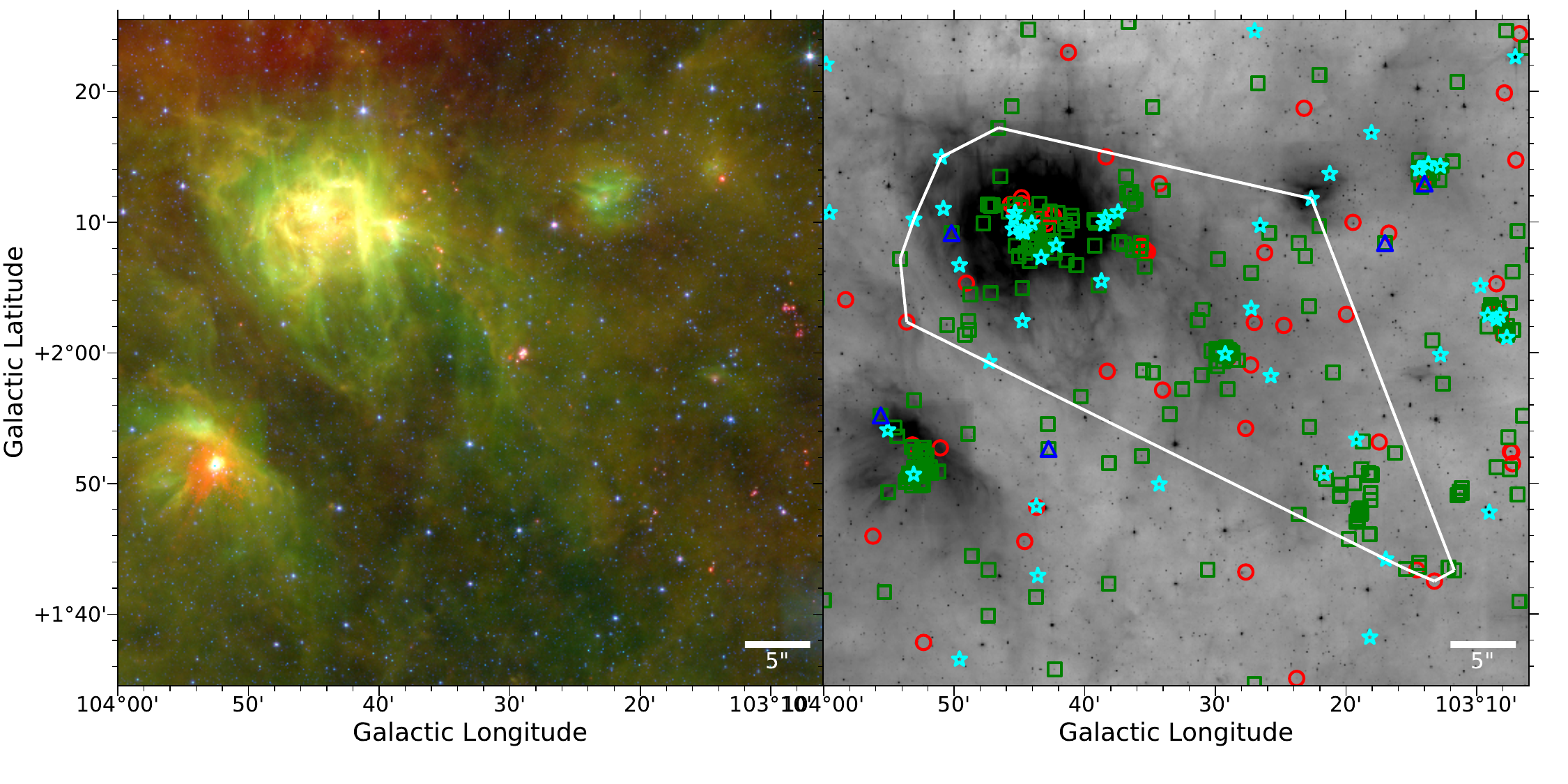}{0.8\textwidth}{}  }
\gridline{\fig{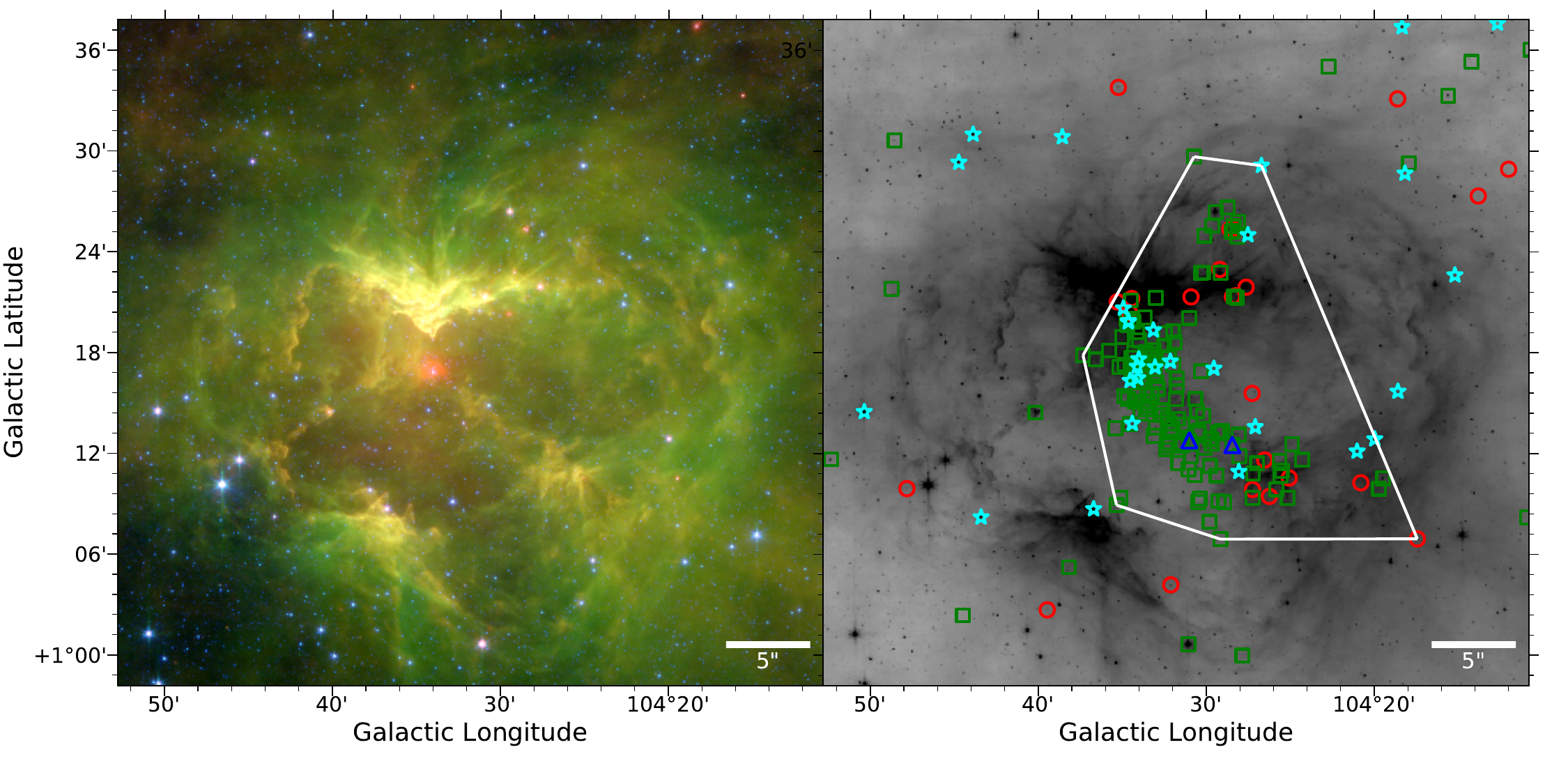}{0.8\textwidth}{}  }
\gridline{\fig{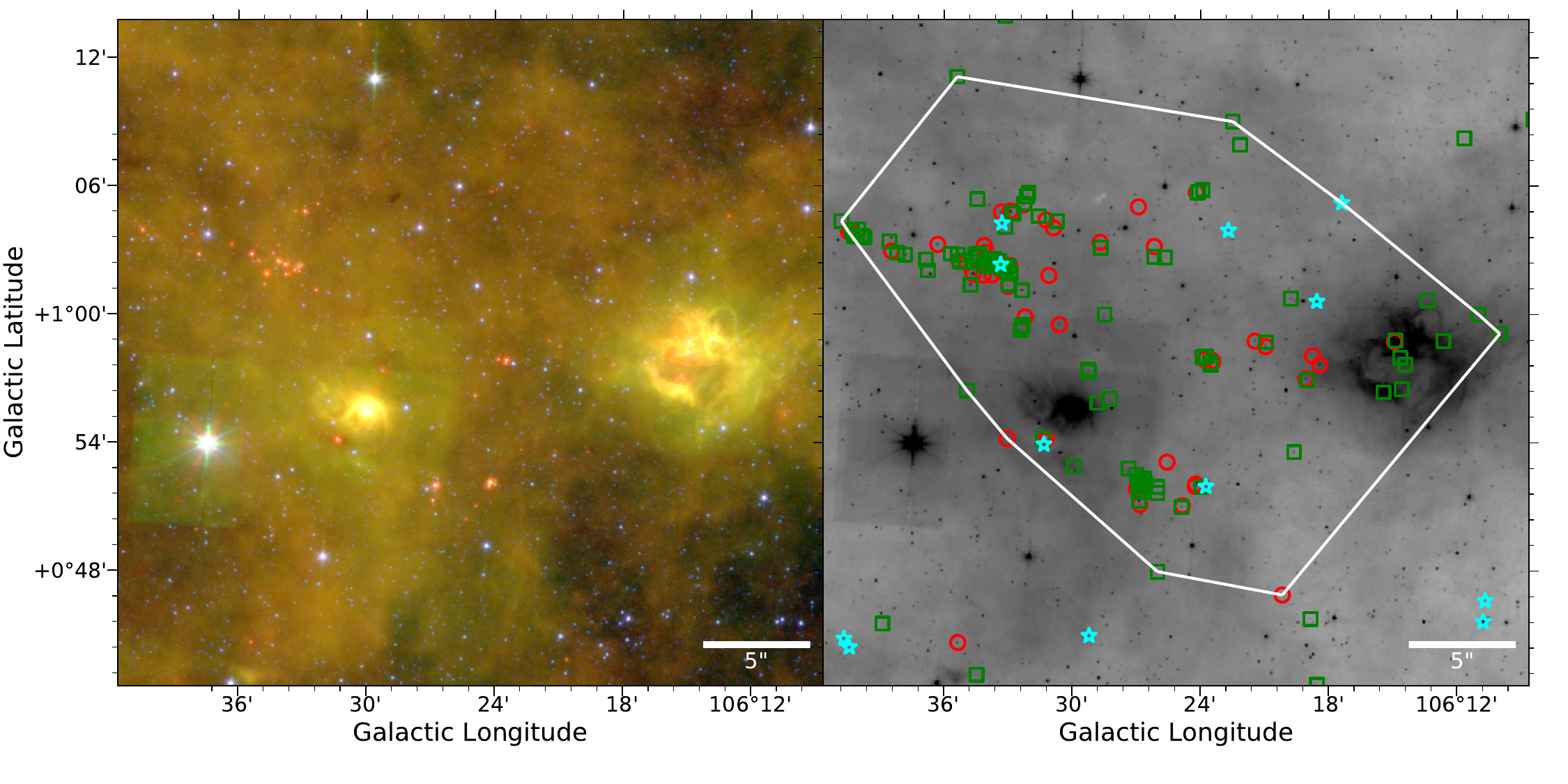}{0.8\textwidth}{}  }
\caption{ Similar to Fig.\ref{fig_grey8_1}, for Clusters \#11, \#30, and \#0 from top to bottom.  }
\protect\label{fig_grey8_2}
\end{figure*}

\begin{figure*}
\gridline{\fig{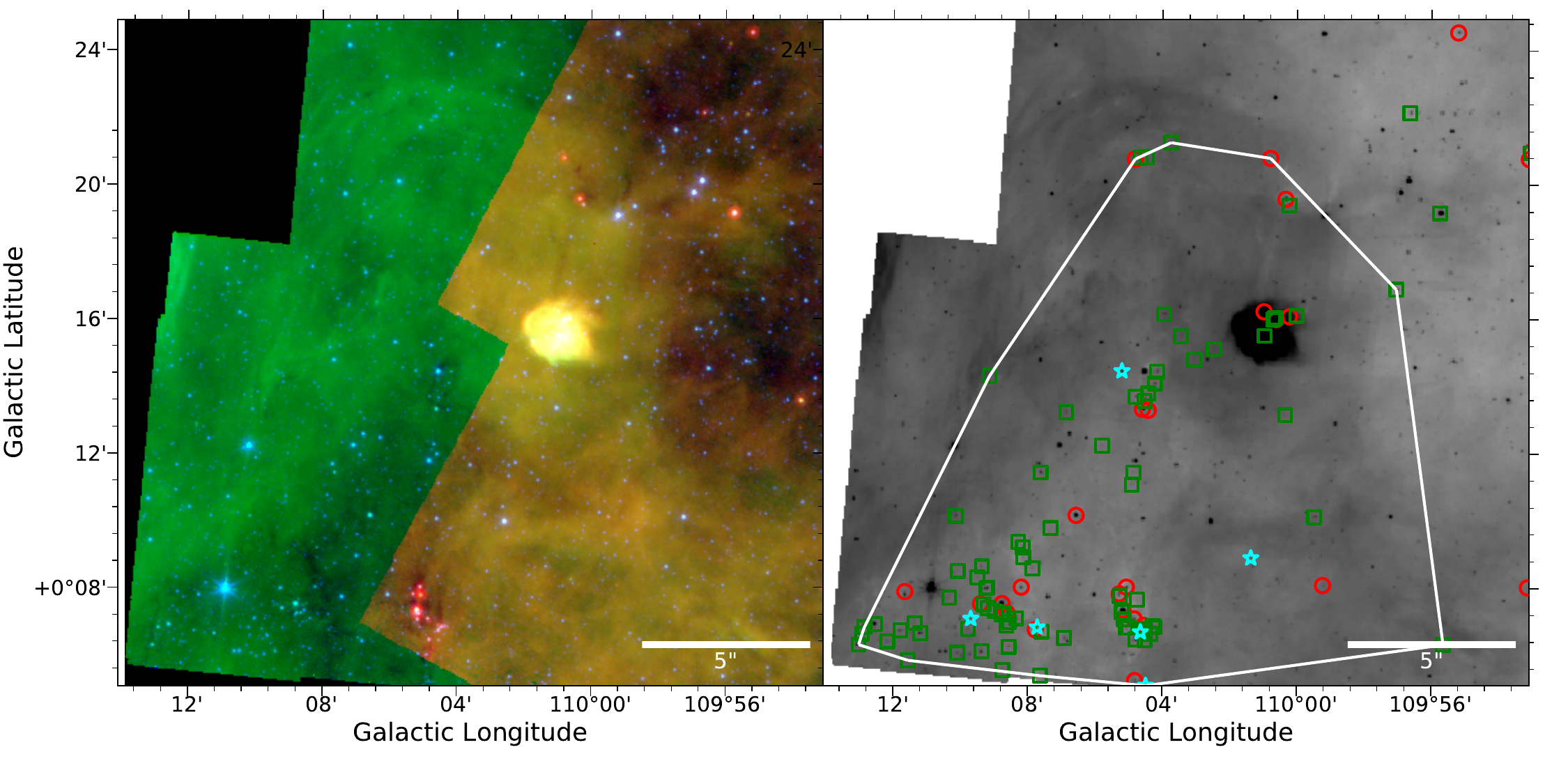}{0.8\textwidth}{}  }
\gridline{\fig{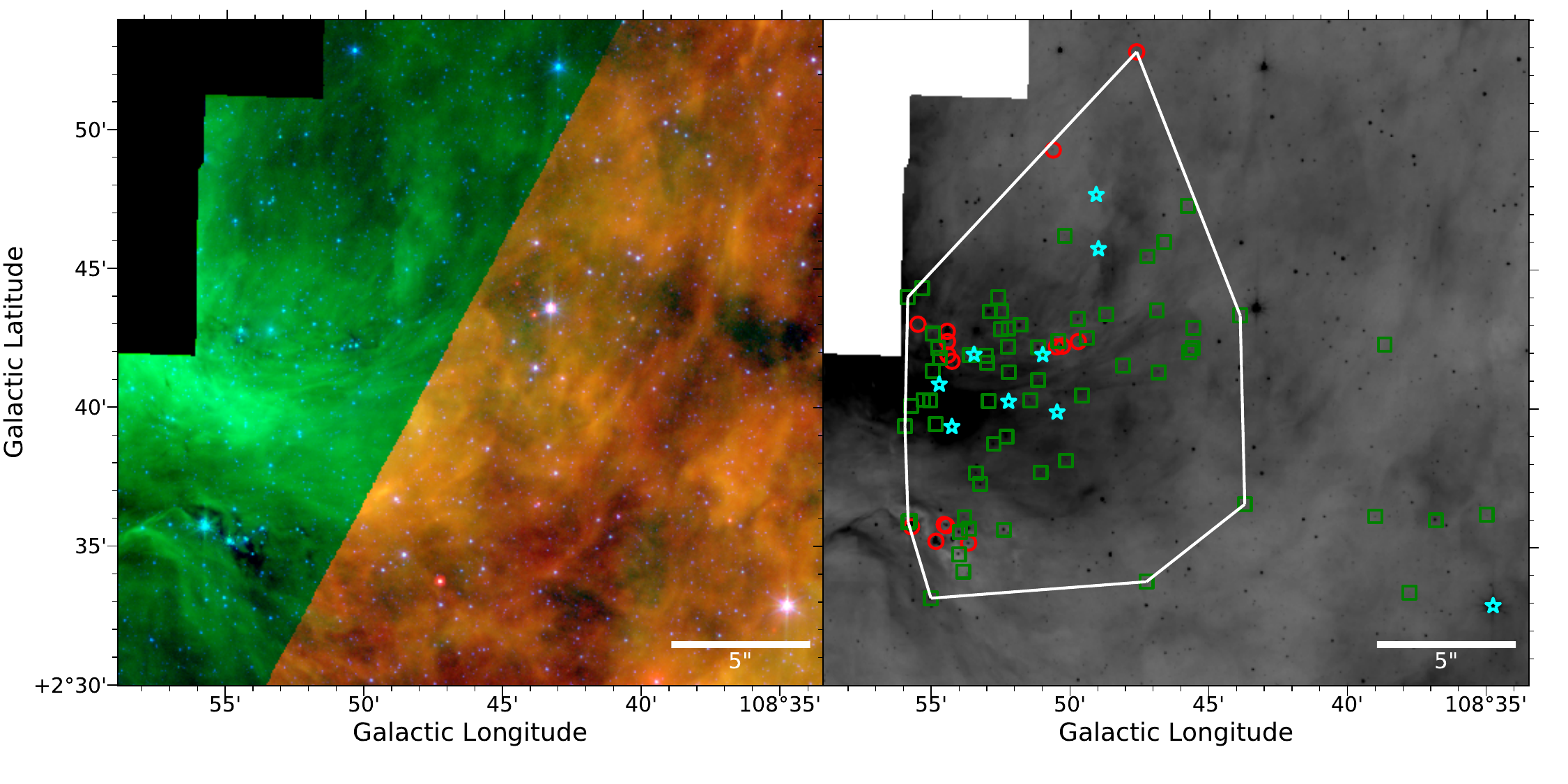}{0.8\textwidth}{}  }
\caption{ Similar to Fig.\ref{fig_grey8_1}, for Clusters \#66 and \#59 from top to bottom.    }
\protect\label{fig_grey8_3}
\end{figure*}

The approximate area of each cluster was quantified by measuring the convex hull of the associated cluster members. 
The convex hull of a set of points in two dimensions is the minimum area polygon that contains those points such that all internal angles 
between adjacent edges are less than 180 degrees. 
Figure~\ref{fig_greycluster} shows the 8.0$\mu$m grayscale image of the SMOG field, with the convex hulls of the clusters overlaid.   
For the eight largest identified clusters, an expanded view of the 8.0$\mu$m 3-color and grayscale is shown in Figures~\ref{fig_grey8_1}, \ref{fig_grey8_2}, and \ref{fig_grey8_3} with the YSOs over-plotted.   

Table~\ref{tab:clusters} lists the properties of the clusters found in the SMOG field; the coordinates of their central point, the number of YSOs and subclusters, {\it WISE} HII 
counterpart, and circular radius based on the separation of the most distant YSOs.  The complete version of the table is available in the electronic version.  The online version 
also includes a full listing of the SIMBAD objects located within the convex hull of each cluster.  These sources are assumed to be associated with the cluster, though no 
attempt has been made to filter the lists or to match them to the YSOs. They are provided as a reference for more in-depth studies.

\subsection{Subclustering}\label{sec_ssc}

For each of the 68 clusters, the spatial distribution of the YSOs was plotted with the convex hull delineating the region; examples of three of the largest clusters 
( Clusters \#0, \#2, \#10) are shown in Figure~\ref{fig_grey8_1}.   As can be seen from this figure, the regions found by the DBSCAN algorithm are not all uniformly spherically 
symmetric, with obvious substructure and asymmetries visible in the three clusters shown.

\begin{figure*}
\epsscale{1.2}
\plotone{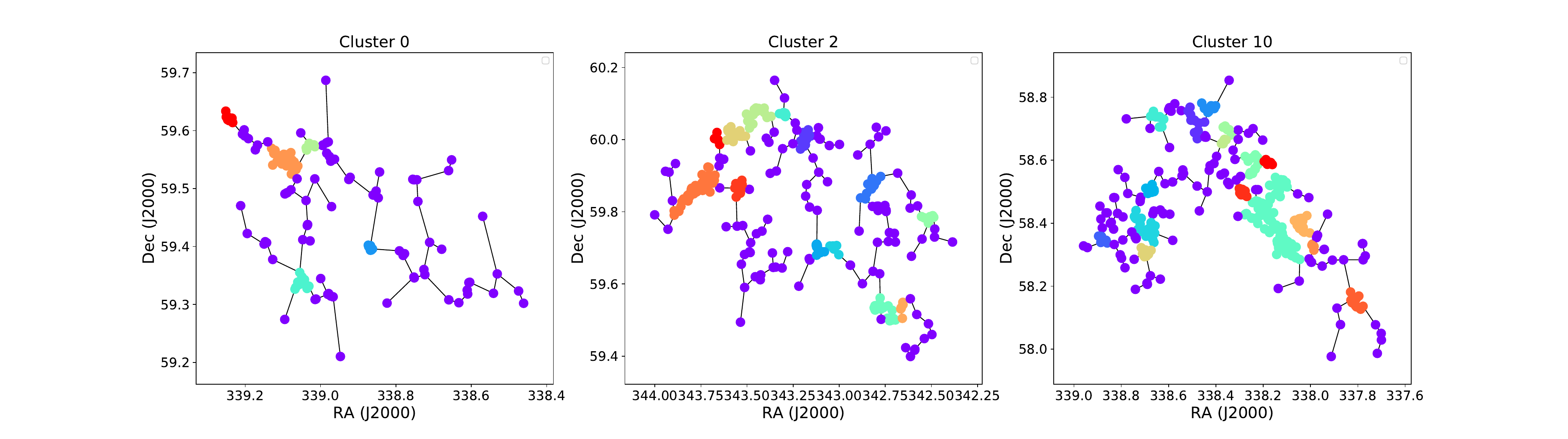}
\caption{ Examples of the subclusters of the YSOs as identified using the Minimum Spanning Tree (MST) method for three of the larger identified clusters are  Cluster 0, Cluster 2, and Cluster 13. 
The black lines show the MST branches.  The purple dots indicate the YSOs which were not assigned to a subcluster.  The various other colors code for the 
identified subclusters with the larger region.  }
\protect\label{fig_3mst}
\end{figure*}

To perform a quantitive search for substructure within the 68 identified clustered regions, a minimum spanning tree (MST) graph of the YSOs was calculated for 
the each of the regions.   This graph connects the YSOs in such as way as to connect each point without loops while minimizing the total 
length of the branches. This was done using the {\it Python} MST clsutering routine written by \citet{vand}.   Subclusters were identified as those points connected by branches with 
lengths less than the characteristic branch length \citep{gut09}.  Figure~\ref{fig_3mst} shows the MST graphs for the Clusters \#0, \#2, and \#10.  The purple dots 
indicate the objects that were found not to be in subclusters, and the brighter colored dots indicate the identified subclusters in each cluster.  

The value of the characteristic branch length was determined separately for each region and is defined as the knee point in the cumulative distribution of branch lengths.
Subclusters were identified in 56 of the 68 of the clusters, or $\sim$82\%.   There were 12 clusters without subclustering, all of which contained 20 members or less.  
For the eight largest clusters with more than 80 members:  Clusters \#0,   \#2,   \#10,  \#11,  \#22,  \#30,  \#59, and \#66, were found to contain 
 6, 13,  18, 8,  15,  3,  3,  and 6 subclusters, respectively.   
 
The result is that the larger the identified clusters/regions the higher the number of subclusters. This is intuitive, and indicates that the larger regions are likely to be comparable 
to the nearby Orion complex in terms of extended structure and multiple formation sites surrounding the massive central core.  Further high-resolution near-IR observations 
of the eight largest regions is underway to better determine their high-mass stellar content and to identify the lower mass and diskless population more completely.  
This will provide a more accurate assessment of the subclustering and whether certain clusters are older/more evolved or if triggering has occurred.  

\begin{figure*}
\epsscale{1.2}
\plotone{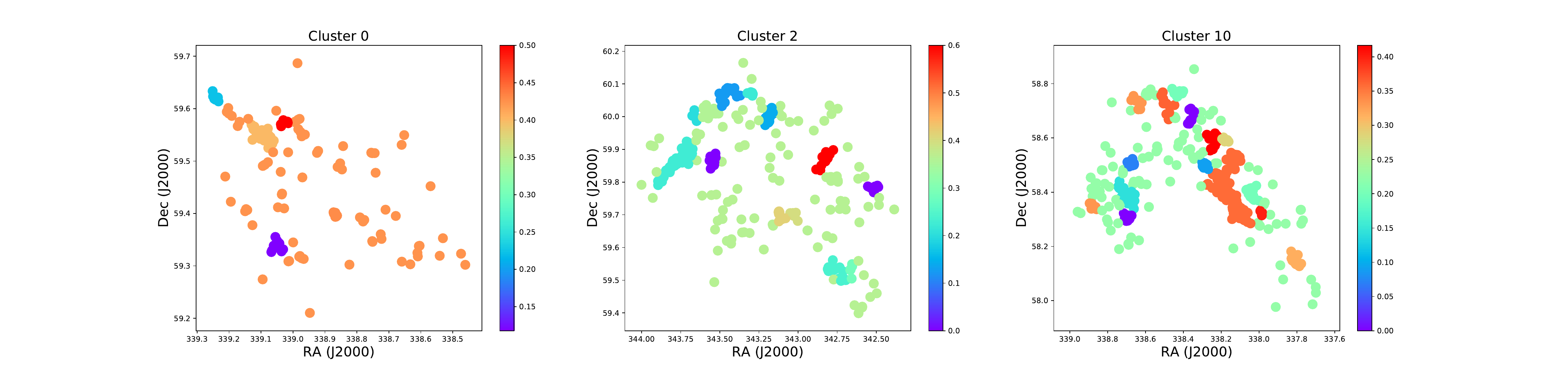}
\caption{Examples of the variation in protostellar fraction, class I/(Class I+Class II), across the subclusters and field with each cluster.   The clusters are the same as in Fig.\ref{fig_3mst}. 
The color bar shows how the color scaling relates to protostellar fraction in each cluster, with red indicating more protostars and blue indicating more pre-main sequence 
YSOs.  It should be noted that the scale is different in each cluster in order to best show variation within its subclusters. }
\protect\label{fig_3psf}
\end{figure*}

Having determined that a number of the larger clusters contained subclusters, an attempt was made to determine whether the protostellar fraction varied within each 
cluster.  The protostellar fraction, as defined here, represents the ratio of the class I to class II YSOs in an area.  
We include only the class I and class II sources in this ratio, as these two classes will be similarly complete based on our data and selection methodology. 
In this case, for each subcluster and for the dispersed population with 
each cluster, the ratio was calculated, and the results plotted for the three Clusters \#0, \#2, and \#10 in Figure~\ref{fig_3psf}.  The dots represent the 
YSOs, and the colors represent the protostellar fraction for each subcluster.  In a number of cases the dispersed population showed a higher ratio than many of the 
subclusters, with knots of mainly class II sources showing very low ratios (as can be seen in Cluster~\#0).  In other cases, as exemplified by Cluster~\#10, the 
subclusters show slightly higher ratios than the dispersed population, with some regions comprised of almost 50\% class I objects.  
There was no trend in protostellar fraction with spatial distribution obviously visible in any of the 68 clusters examined. In the majority of cases, the variation in protostellar fraction was small, 
a factor of $\sim$2-3 across the region.  In a few cases, a factor of $\sim$5-10 was present in regions where small knots of both class I and II sources were present.   
Though incomplete without the inclusion of a confirmed population of class III young stars to provide accurate disk fractions as a measure of cluster age, this study 
nevertheless provides a useful indication of variation in age and possibly environment across the identified regions.

\section{SED model fitting}\label{sedfitter}

The SED has previously been used to broadly classify each of the YSOs into 
an evolutionary stage based on its slope across the IRAC bands.  
The Python SEDFitter package of \citet{robitaille} uses a sample grid of YSO model SEDs with varying age, mass, inclination etc., to compare to the input photometry, with the scale factor $S$ 
(dependent on source distance and luminosity) and the extinction $A_V$ as free parameters.  The code returns a sample of best fit models and their 
associated parameters.  

An initial run was performed for all identified YSOs, separated into clusters and the field population, where the distance was allowed to range from 1-8~kpc 
and the extinction to range from 0-40~$A_V$.   It was found that within each cluster the range in distances for the best fit models covered the entire provided 
distance range, and that therefore the results could not be used in any meaningful way.  

It was therefore decided that only those clusters with reliable estimates for their distances would be included in the SED Fitter modeling.  
Reliable distances were found for 12 of the 68 clusters in the SMOG field.   The {\it WISE} HII survey was consulted for known objects in the field \citep{anderson}.  
These were overplotted on the clusters and radial distance matching as well as spatial coincidence of the convex hulls with the HII region determined.  
A final visual assessment indicated an association between HII regions with distance estimates and 12 of the clusters.  The associated HII regions are listed in Table~\ref{tab:clusters}.    
The clusters with identified distances were: Clusters \#0, \#2, \#10, \#11, \#25, \#30, \#32, \#39, \#41, \#60, \#64, \#66.  
The distances ranged from 2.5~kpc to 7.8~kpc, and are listed in Table~\ref{tab:wise_distances}. 
\citet{cooper} provide distances to three of these clusters from their spectral survey of high mass YSOs:  Clusters \#10, \#39, and \#41.  
The distances to Clusters \#39 and \#41 matched exactly, their distance to Cluster \#10 was 2.9~kpc, whereas the WISE distance was 5.7~kpc.  
We elected to use the WISE distance in order to have a consistent distance metric for all the modeled clusters.  
These distances indicate that the SFRs are distributed over a wide section of the outer Galaxy, and are located close to both the Perseus Arm and the Outer Spiral Arm, 
as well as falling at intermediate distances indicating that they may be in the inter-arm region.    

The second release of the {\it GAIA} catalog was also searched for matches between known YSOs.  
For two clusters, a previously identified YSO within the convex hull was matched to a {\it GAIA} source within an estimated distance of 1$\arcsec$.  
The {\it GAIA} parallaxes did not provide distances within the ranges of the spiral arms.  
Therefore, considering that there was a high likelihood that the matches were erroneous, we did not include them in the subsequent analysis.  

SEDFitter was rerun using the distance range fixed to the known distance, while the $A_V$ was still allowed to vary from 0-40~$A_V$, for each cluster 
separately.   Not all of the YSOs could be fit, those lacking photometry across a sufficient number of bands were not fit.  
Table~\ref{tab:SEDfit} lists a sample of the results of the SEDFitter routine for the YSOs in the 12 clusters for which distances were known.  A number of parameters
are presented for each model fit, including the best fit to the $\chi$-squared, object mass, disk mass, age, $A_V$, central temperature, disk accretion rate, etc.  
The upper and lower limit model fit parameters are also supplied in each case.  The full data table is available online in electronic format.  

\begin{figure*}
\epsscale{1.2}
\plotone{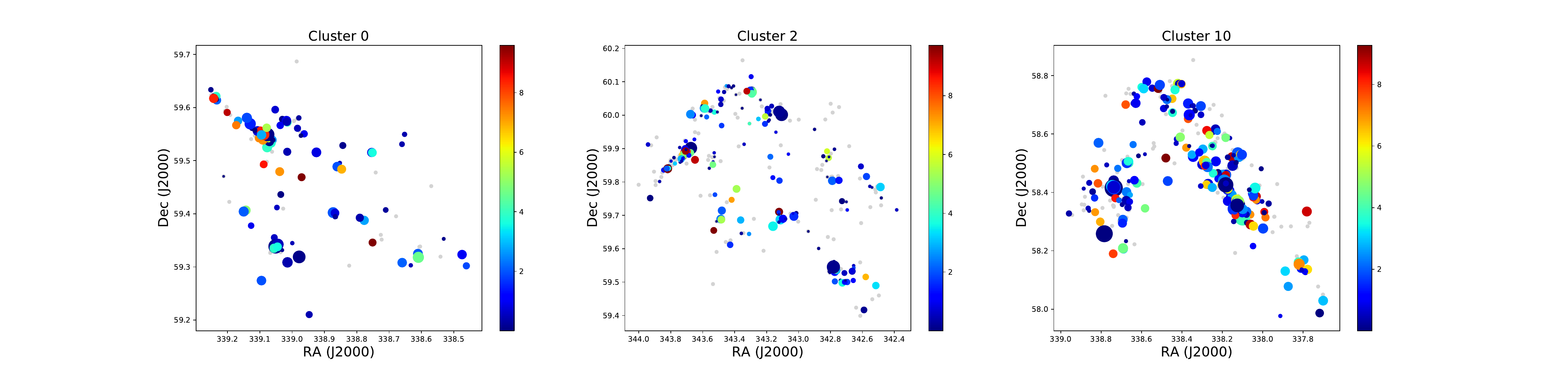}
\caption{ Examples of the SEDFitter results for spatial distributions of ages and masses within each cluster. The clusters are the same as in Fig.\ref{fig_3mst}. 
The relative sizes of each symbol indicate how massive the YSO is with respect to the most massive object in that cluster.   
The color bar shows how the color scaling relates to the calculated age of the YSO, with the bluer objects being younger, and the redder objects being older.  }
\protect\label{fig_3sedam}
\end{figure*}

The individual ages and masses derived from the model fits may not be entirely reliable and will not be discussed here. However, cumulatively by cluster, they 
can provide an insight into the relative ages and masses of the YSOs, and thus provide an estimate of the age and mass range of the cluster as a whole.   
Figure~\ref{fig_3sedam} shows the age and relative mass of the stars in three clusters by their spatial distribution.  The size of the circles indicates the mass 
of the star relative to the most massive object in that cluster.  The color indicates the age, the range in ages in Myrs is shown by the colorbar for each plot.    The modeled 
YSOs are predominantly very young, being $\sim$1-3~Myrs old, with some ranging up to 10~Myrs.  
The more massive objects tend to be younger, however, this may in part be due to a bais in the selected grid of models \citep{robitaille2006}.  
None of the 12 clusters showed any  indication of a trend with age across the subclusters or dispersed population. Though some subclusters appeared to 
be slightly younger on average, these tended to be the ones containing class I YSOs, which are by definition younger.  None of the clusters showed any geometric trends in age 
distribution across the cluster, an indication of possible triggering or sequential star formation episodes.  

\begin{figure*}
\plottwo{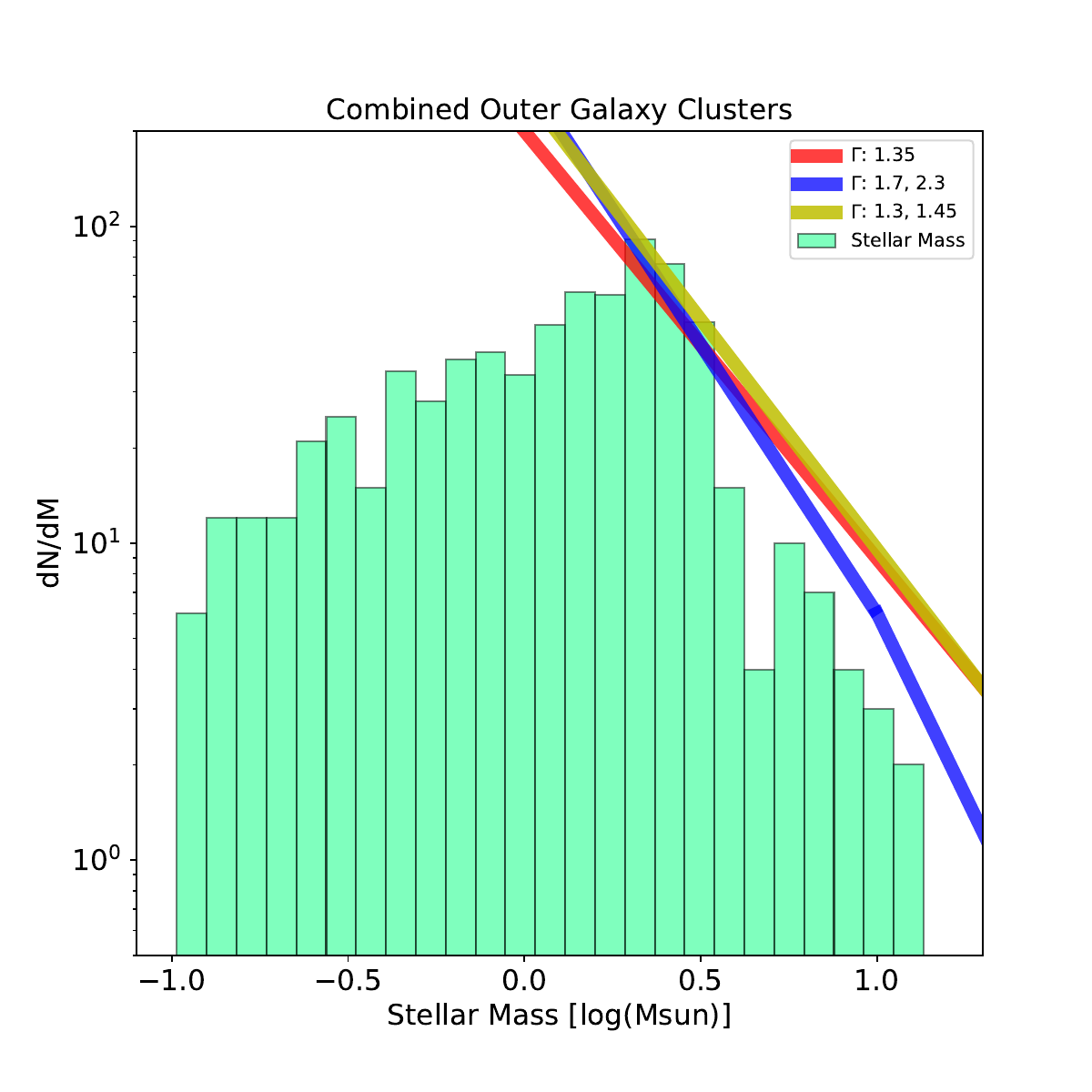}{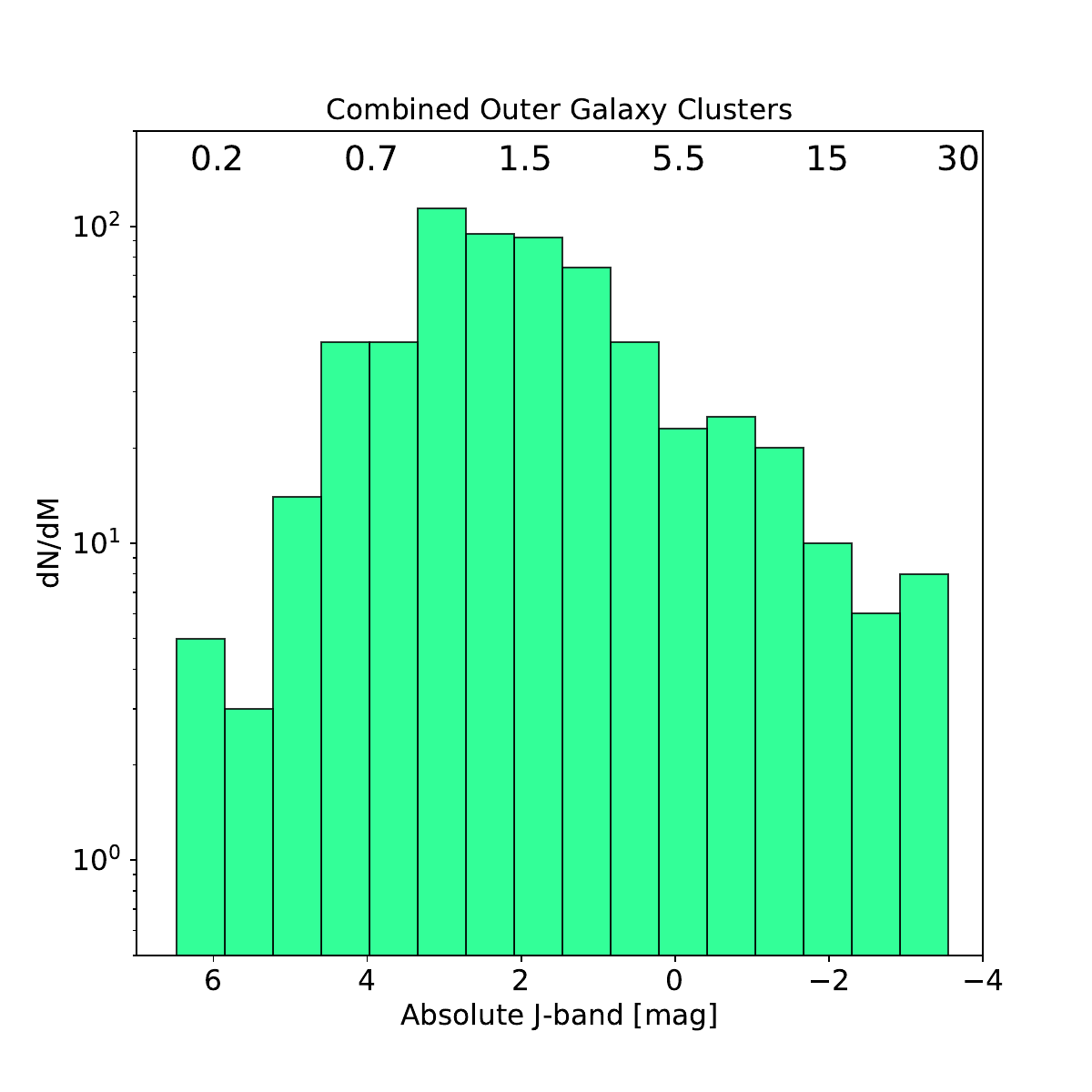}
\caption{ Examination of the Initial Mass Function for the outer galactic regions identified in the SMOG field.  
{\it Left:} The IMF as determined by combining the SEDFitter calculated masses for the 12 clusters with distance estimates.  The green histogram shows the 
best fit model masses for the YSO.  The data show reasonable correlation to both the Salpeter slope of the IMF (m$\sim$-1.35) and more recent estimates (m$\sim$-2.7, m$\sim$-2.3) for a broken power law fit.  
{\it Right:}  A photometric version of the initial mass diagram, using the dereddened absolute J-band magnitude of the YSOs in the 12 clusters.  The approximate stellar masses 
are taken from the MIST isochrones at 5~Myrs with $v/v_{crit} = 0.4$ for the 2MASS passbands \protect\citep{mist}.  
}
\protect\label{fig_imf}
\end{figure*}

Figure~\ref{fig_imf} (left) plots the modeled IMF combining all of the YSOs with estimated stellar masses obtained via fitting of their SEDs from each cluster.  
The green histogram shows the best fit model masses, where the model reduced $\chi$-squared was $\le$ 3.  
The universality of the IMF is still debated; the IMF for high mass stars ($M_{\odot} \ge 1$) can be described as $\Psi(m) \propto m^{-\Gamma}$ \citep{bastian}.   
The red line plots the power law slope with $\Gamma = 1.35$, roughly the value of the Salpeter slope \citep{salpeter}.  The blue line shows the broken power law fit with 
slopes of $\Gamma_1 = -1.7$ and $\Gamma_2 = -2.3$ as described by \citet{miller}.   
A modified version of the Kroupa IMF with three distinct values of $\Gamma$ for the low, solar, and high mass regions \citep{kroupa} is shown in yellow, as presented by 
\citet{weisz} who used the Panchromatic Hubble Andromeda Treasury (PHAT) program to fit the high-mass end of the IMF ($>2M_{\odot}$) of young, intermediate mass clusters 
in Andromeda to derive the value of the IMF slope of $\Gamma = 1.45$ above two solar masses.  
The righthand plot shows the extinction-corrected absolute $J$-band magnitudes of all the YSOs in the clusters with good photometry.  The extinctions, $A_K$, are obtained from the photometry, all 
objects in a cluster are presumed to lie at the same distance, and the $J$-band magnitudes are taken from 2MASS and UKIDSS when available.   Estimates of the stellar masses corresponding 
to each magnitude were obtained from the $MIST$ website of isochrone packaged model grids for 5~Myr~old stars with $v/v_{crit}=0.0$ \citep{mist}.     
A recent review of the slope of the high-mass end of the IMF by \citet{parravano} found that steeper values in the range $\Gamma = 1.7-2.1$, as reported in 
e.g. \citet{dawson}, could be accounted for by the effects of local environment, our position in an inter-arm region, and the diffusion of the lower mass stellar population 
over their lifetimes, and that the Salpeter value of $\Gamma = 1.35$ could thus be considered universal.  
A linear regression fit to the histogram for the higher mass end of the distribution from $\approx$ 2-10 M$_{\odot}$ returns a fit with a slope of  $\Gamma = 2.38 \pm 0.20$ 
for the combined clusters;  a similar fit from $\approx$ 4-10 M$_{\odot}$ returns a fit with a slope of  $\Gamma = 1.77 \pm 0.25$ for the combined clusters.   
These $\Gamma$-values are broadly consistent with the higher slopes reported in the literature. 
We therefore conclude that slope of the IMF in the star forming regions so far assessed in the SMOG field is consistent with the universal Salpeter IMF and the inner Milky Way galaxy.

\section{Eight Largest Identified Regions}

The $DBSCAN$ method identified eight regions of ongoing star formation with more than 80 identified members from the combined IR catalogs. 
These regions were: Clusters  \#0,   \#2,   \#10,  \#11,  \#22,  \#30,  \#59, and \#66, with 155, 318, 428, 198, 221, 174, 80, and 104 members, respectively.  
Each of these regions is associated with a region of bright emission in the 8.0$\mu$m image, generally associated with emission from dust grains heated 
by the stellar irradiation of massive stars.  Each of the regions shows evidence of subclustering within the convex hull.  
An expanded view of the 8.0$\mu$m grayscale around each of the eight clusters is shown in Figs~\ref{fig_grey8_1}, \ref{fig_grey8_2}, and \ref{fig_grey8_3} with the YSOs and convex-hulls over-plotted, alongside 
a similarly scaled three color image of the cluster at 4.5, 8.0, and 24$\mu$m.   

For six of the eight clusters (all but Clusters \#22 and \#59) estimates of their distances have been obtained by matching to the {\it WISE} HII survey catalog.  
This allows us to estimate their two dimensional physical spatial extent.  We used two values for the cluster radius: the estimated circular radius and the 
estimated effective hull radius of each cluster.  The calculated spatial scales of the seven clusters for [$R_{circ}$, $R_{eff-hull}$] were approximately: 
[90, 87], [71, 48],  [129, 104], [48, 45], [34, 45], [61, 70] in parsecs, for Clusters \#0,   \#2,  \#10,  \#11,  \#30, and \#66, respectively.  
These size scales are comparable to massive star forming complexes such as Orion, with a span of $\sim$80~parsecs across the Orion A and B clouds \citep{megeath}, 
and RCW~38 with a span of $\sim$20~parsecs across the immediate cloud complex surrounding IRS~2 \citep{win11}.  

The most massive YSOs identified by SEDFitter in the six clusters with distance estimates were:  
9,5, 10.5, 11.1, 7.0, 5.7, and 13.6~M$_{\odot}$, for Clusters \#0,   \#2,  \#10,  \#11,  \#30, and \#66, respectively. 
These masses correspond to spectral types from early B to late O-type stars.  

Further, by applying the modified Kroupa power law distribution for the IMF \citep{weisz} discussed in Sec.~\ref{sedfitter} we can estimate the total population
above $\approx$0.5 solar masses and compare them to similar regions in the inner Galaxy.  We made the assumption that the bins for stars with 2-5~M$_{\odot}$ were 
complete and used this value of $N$ to calculate the value of $\epsilon_0$ for the two mass regimes:
\begin{equation}
N = (\epsilon_0 / \Gamma) * [ M_1^{-\Gamma} - M_2^{-\Gamma} ]
\end{equation}
, where $M_1 = 2_{\odot}$ and $M_2 = 5_{\odot}$. 
The calculated values of $\epsilon_0$ were then used to obtain estimates of cluster membership for 0.5-1~M$_{\odot}$ and 1-100~M$_{\odot}$ and then summed to give 
total population estimate for the cluster above $\approx$0.5 solar masses.  The estimated memberships were: 
460, 265, 1320, 205, 195, and 285 members for Clusters \#0,   \#2,  \#10,  \#11,  \#30, and \#66, respectively. 
These were compared to values of 1000-1500 for Orion \citep[estimated from][]{megeath} and $\sim$650 for RCW~38 \citep{win11}.  
The Orion cluster lies at a distance of 414~pc and we can thus identify young stellar members to well below the hydrogen burning limit.   
The RCW~38 complex lies at a distance of 1.7~kpc, and thus may more closely reflect the detection rate for intermediate mass stars with IRAC.   
All clusters show similar or lower estimates to the RCW~38 complex, which has been estimated to have a total stellar population of 10$^4$ YSOs.  
This would imply that at least six of the eight regions are medium-sized star-forming complexes similar in extent and stellar density to those found in the 
inner Galaxy.

\section{WISE Comparison with IRAC}\label{wise_compare}

\begin{figure*}
\epsscale{1.2}
\plotone{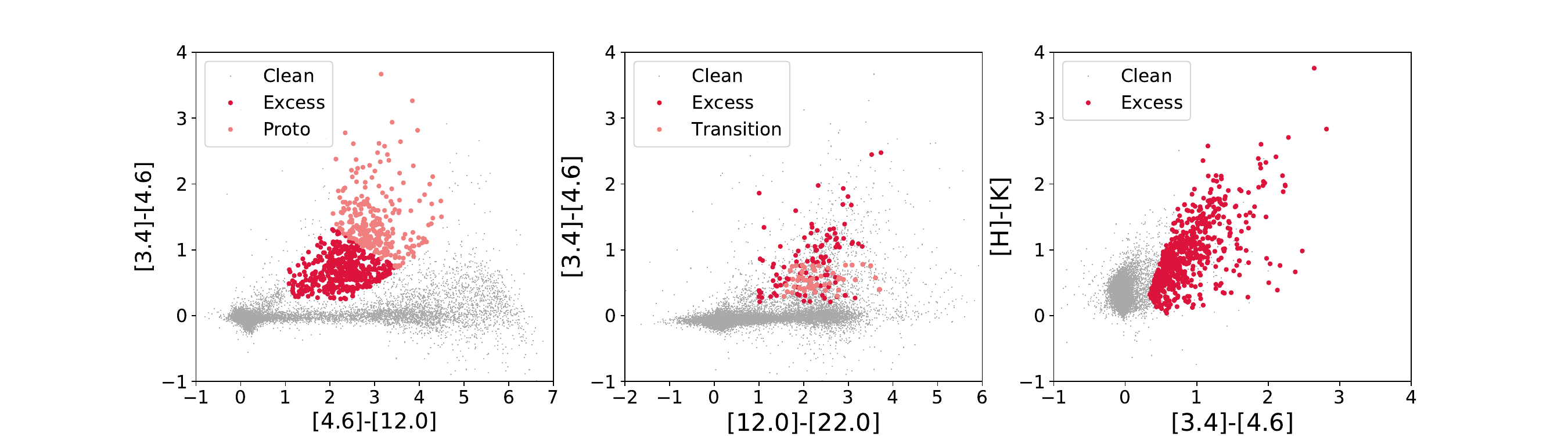}
\caption{ Color-color and color-magnitude diagrams showing the {\it WISE} YSO selection criteria.  The red/pink dots in each plot indicate the selected YSOs.  
The gray points indicate the cleaned {\it WISE} photometric catalog.  
{\it Left:} [3.4 - 4.6] v [4.6 - 12.0]:  3-band {\it WISE} selection criteria, useful for objects lacking 22$\mu$m photometry, most similar to IRAC selection criteria.  
{\it Center:}  [3.4 - 4.6] v [12.0 - 22.0]: Sources with full four band {\it WISE} photometry, useful for selecting candidate transition disk objects.  
{\it Right:} [H - K] v [3.4 - 4.6]: Combined 2MASS and {\it WISE} bands 1 and 2, for sources lacking {\it WISE} longer wavelength detections.  
}
\protect\label{fig_wise}
\end{figure*}

The {\it WISE} catalog of the SMOG field contained 581,705 point sources, or $\sim$20\% the number of detections in the SMOG catalog.  
A search for YSOs was performed using the four band {\it WISE} data combined with 2MASS near-IR photometry; a detailed discussion is given in Appendix~\ref{wise_ysos}.   
After the removal of spurious sources and contaminants, there were 16,689 point sources remaining in the catalog.   
Figure~\ref{fig_wise} shows three of the plots used to select YSOs, with the gray points indicating the cleaned catalog, with the red points indicating the 
selected YSO candidates.  
Of the 16,689 point sources, a total of 931 were identified as YSOs.  This number is roughly one-fifth (931/4648 or 20\%) of the YSOs detected using the IRAC methods.
The evolutionary classifications of these YSOs were determined to be: 271 class I, 641 class II, and 19 candidate transition disks.  

The {\it WISE} full and cleaned catalogs were matched to the SMOG full and cleaned catalogs with both a 1$\arcsec$ and 1.5$\arcsec$ matching radius.  
The catalogs of detected YSOs were also compared.  The numbers of sources matched between catalogs are reported in Table~\ref{wise_irac_match}.   
Of the 581,705 sources in the full catalog, 456,530 (78\%) were matched to the full SMOG catalog.  The percentage is higher at 89\% for the two cleaned 
catalogs, suggesting that a number of spurious detections have been removed from the {\it WISE} catalog.  There was an overlap of 68\% 
between the two YSO catalogs, indicating that 32\% of the {\it WISE} YSOs may be misidentifications or contaminants.   

Table~\ref{tab:WISEphot} lists column information for the data table of {\it WISE} identified YSOs, including the {\it WISE} photometry, IRAC identifier, and whether or 
not the object was selected as a YSO by IRAC.   The full data table is available online in electronic format.   
   
We compared the IRAC bands 1 \& 2 (3.6 \& 4.5$\mu$m) photometry to {\it WISE} bands 1 \& 2 (3.4 \& 4.5$\mu$m) for those 
sources matched at 1$\arcsec$ in the cleaned catalogs.  The expected linear relationship between the magnitudes was found with increasing scatter towards 
fainter magnitudes, which is expected due to source confusion and differences in background subtraction.  The histograms of the differences in magnitude 
for all matched sources and the YSOs were also examined.   In all instances, the peak of distribution is very close to zero, indicating relatively good photometric precision 
between the two instruments.  The median values for [I1 - W1] were -0.013, -0.127, -0.153~mag for all sources, class I, and class II, respectively.  
The median values for [I2 - W2] were -0.019, 0.237, 0.049~mag for all sources, class I, and class II, respectively.  

\begin{figure*}
\epsscale{1.2}
\plotone{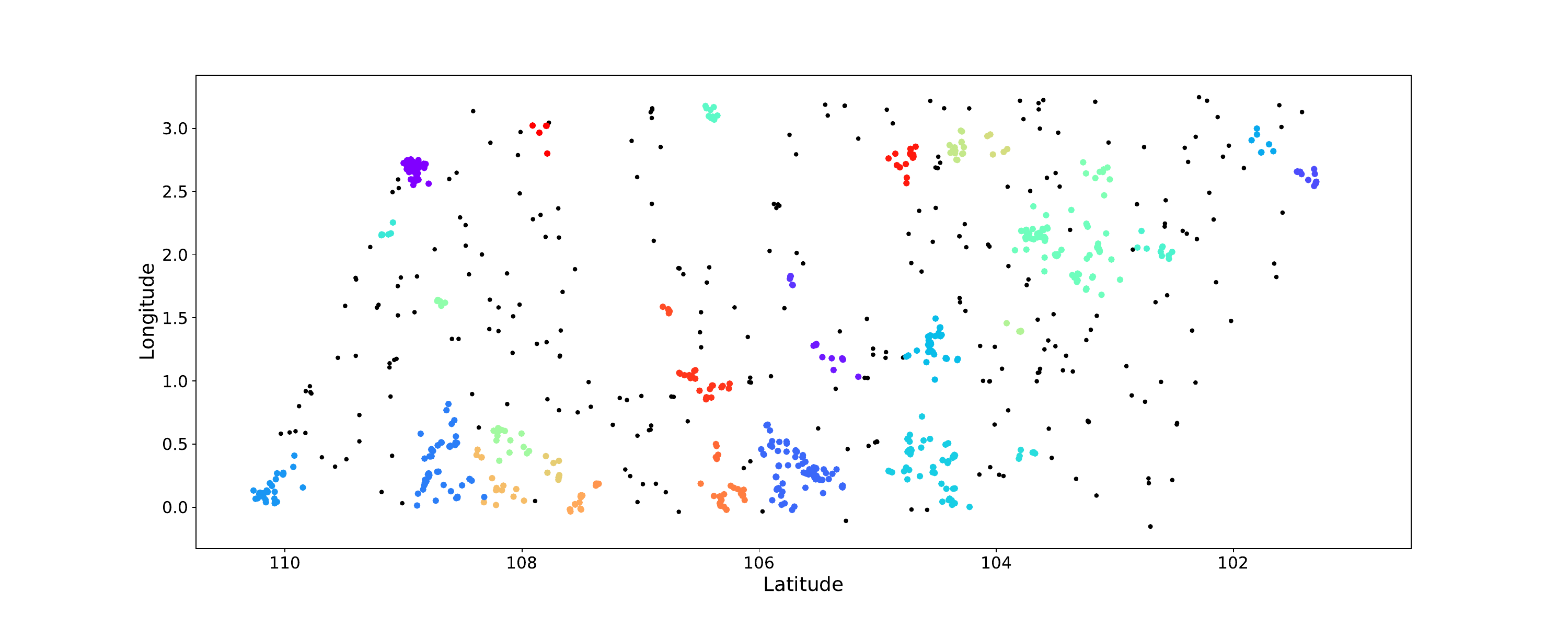}
\caption{  Spatial distribution of the {\it WISE} identified YSOs, showing the clusters identified by the DBScan method.  The sources in each cluster are color-coded.  
The black dots represent those YSOs not identified as belonging to a cluster.  The identified clusters are similar to the larger clusters identified with IRAC.  
}
\protect\label{fig_wdb}
\end{figure*}

\begin{figure*}
\epsscale{1.2}
\plotone{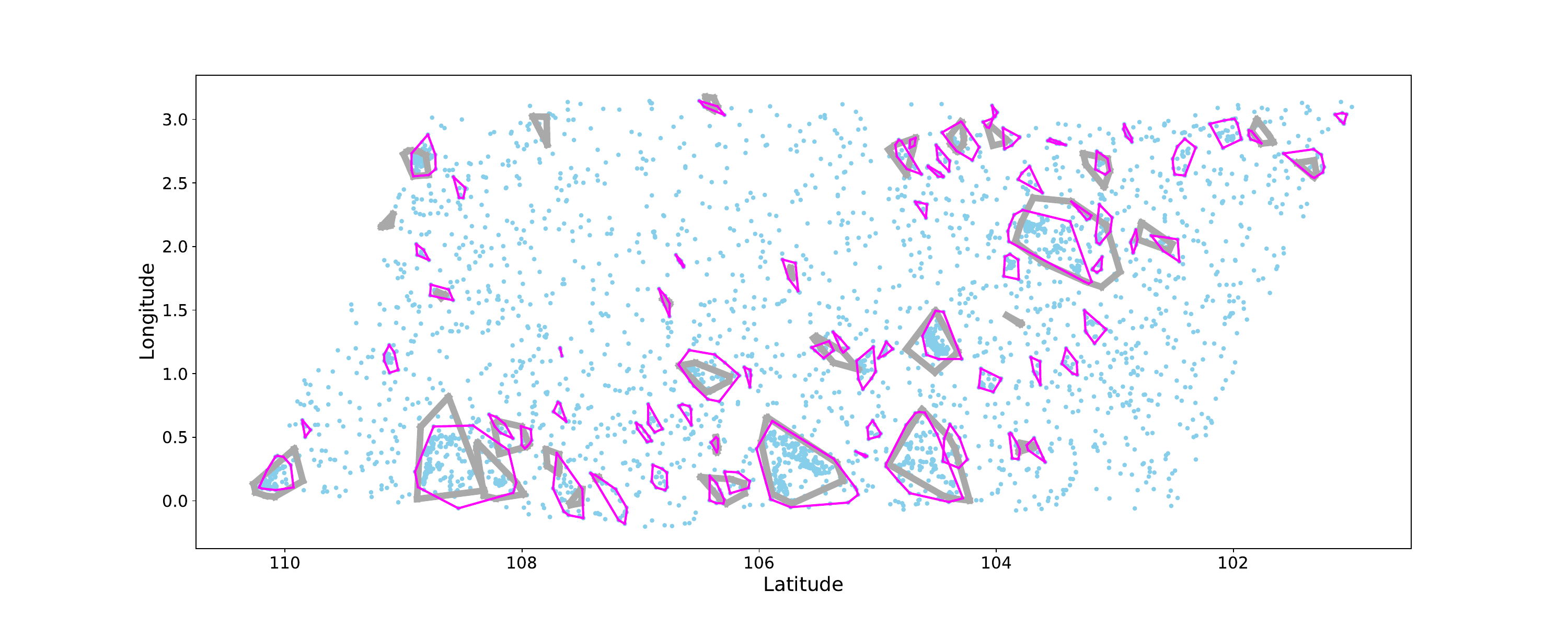}
\caption{  Spatial distribution of the {\it WISE} identified clusters overlaid with the IRAC identified clusters, as shown by their convex hulls.  The {\it WISE} clusters (gray) are 
broadly consistent with the IRAC clusters (magenta).    The blue dots represent the IRAC identified YSOs, showing that there are 
sources detected in the within the {\it WISE} clusters.  
}
\protect\label{fig_sdcomp}
\end{figure*}

The spatial distribution of the {\it WISE} identified YSOs was examined in the same manner as with the large scale clustering of the IRAC 
sources, using the DBSCAN algorithm.  In this case, the separation length was set to $\epsilon = 0.2$ based on the turn-off in the cumulative 
distribution of nearest neighbors.  A total of 31 clusters were identifed, as shown in Figure~\ref{fig_wdb}.  These clusters are broadly consistent 
with the locations of the IRAC identified clusters, though a number are formed from agglomerations of smaller regions or do not extend to the same 
area as the IRAC-identified clusters due to the detection of fewer, more distributed YSOs.  
Figure~\ref{fig_sdcomp} shows an overlay of the spatial distribution of both the {\it WISE} clusters and the IRAC clusters, as outlined by the convex hulls of 
the sources in each cluster, with the blue dots representing the IRAC identified YSOs.  From this figure, we suggest that the {\it WISE} data is, for the greater part, locating the same 
star formation sites as the IRAC data but with a lower spatial resolution or fewer YSOs, and that further studies with {\it WISE} alone would be sufficient to locate areas of 
massive star formation especially.    

An investigation was undertaken to see whether there would be an advantage to combining {\it Spitzer} warm mission observations with the longer wavelength 
WISE data to extend the color-space available to identify disks. 
 A similar comparison was carried out by \citet{sewilo} in CMa-l224, who compared their {\it Spitzer} GLIMPSE~360 and AllWISE YSO selection to the {\it WISE}-only 
selection of \citet{fischer} and found that 20/93 YSOs had been misidentified in the WISE-only sample. 
 For those {\it WISE} detections with matching IRAC photometry, the two shortest {\it WISE} bands 
were replaced by the two shortest IRAC bands and the {\it WISE} selection criteria were run again to ascertain whether a similar number of YSOs would be retrieved.   
It should be noted that the number of objects now available to be searched decreased from 16,689 to 15,199 sources. Using the same selection criteria, the number 
of YSOs detected on the [I1 - I2] v [I2 - W3] diagram dropped from  271 class I and 433 class II objects, to 97 class I and  397 class II objects.   
This represents a loss of 210 sources, or 29\% of the YSOs selected with {\it WISE} only.  Given the close matching in photometry between the IRAC and {\it WISE} data and 
the spatial positions of the unmatched YSOs in the cores of clusters and in close proximity to the borders of the SMOG field, the decrease in detections is attributed to the 1,490 
sources that were not matched between the catalogs, likely due to sources being at the edge of the IRAC field of view and to crowding in the more densely populated star forming regions.

\section{Summary}\label{summ}

We have undertaken a  study of the 21 square degree SMOG field which was observed with the IRAC and MIPS instruments.  
We combined the {\it Spitzer} data with 2MASS and UKIDSS near-IR photometry.  We also compared our results to the {\it WISE} catalog of the field to assess the usefulness of 
merging these two datasets when selecting YSOs.   

\begin{itemize}

\item We identify 4648 YSOs with IR-excess emission, of these 1197 were class I, 2632 class II, 819 class III, and 45 were candidate transition disk objects. 

\item Using the DBSCAN algorithm, we identify 68 clusters with $\epsilon_{0}=0.11$ and $N_{min}=8$ YSOs across the SMOG field, all of which had 8 or more members.  
The ratio of YSOs identified as members of clusters was 2873/4648, or 62\%.  
The eight largest clusters:  Clusters \#0,   \#2,   \#10,  \#11,  \#22,  \#30,  \#59, and \#66, had 155, 318, 428, 198, 221, 174, 80, and 104 members, respectively.  

\item Subclusters within the 68 regions were found using the MST of the YSOs with a characteristic branch length estimated for each region.   Of the 68 clusters, 
56 or $\sim$82\% were found to have subclusters.  The remaining 12 regions had 20 members or less.  
The eight largest clusters had  6, 13,  18, 8,  15,  3,  3,  and 6 subclusters apiece.  

\item For the 12 clusters with known distances, the SEDs of the YSOs were fitted using the SEDFitter routine.  From the modeled masses, the IMF was constructed for the known 
clusters in the field.  The slope of the combined IMF was found to be $\Gamma = 2.38 \pm 0.20$ above 2~M$_{\odot}$ and $\Gamma = 1.77 \pm 0.25$ above 4~M$_{\odot}$ . 
These values are consistent with those obtained in other inner galaxy high-mass SFRs, and are likely also consistent with a universal Salpeter IMF.  

\item  With the {\it WISE} catalog we identified 931 YSOs across the SMOG field. The differences in the {\it WISE} photometry as compared to IRAC were found to be negligible when compared 
to the infrared excess emission used to select the YSOs.  Even so, compared to the number of YSOs detected with IRAC,  931/4648 or 20\%, we find that the {\it WISE} 
resolution hampers the detection of individual cluster members.  However, the 32 clusters detected compared favorably in location to the larger IRAC detected clusters, though some 
individual clusters identified with IRAC are merged in WISE.

\end{itemize}

\acknowledgments

We would like to thank the anonymous referee for their helpful comments on the manuscript.  
This work is based on observations made with the {\it Spitzer} Space Telescope (PID~30574), which is operated by the Jet Propulsion Laboratory, California Institute of Technology 
under NASA contract 1407.   E. Winston, J. Hora, \& V. Tolls gratefully acknowledge funding support for this work from NASA ADAP grant NNX16AF37G.
This publication makes use of data products from the Two Micron All Sky Survey, which is a joint project of the University of Massachusetts and the Infrared Processing and 
Analysis Center/California Institute of Technology, funded by the National Aeronautics and Space Administration and the National Science Foundation. 
Support for the IRAC instrument was provided by NASA through contract 960541 issued by JPL. 
This publication makes use of data products from the Wide-field Infrared Survey Explorer, which is a joint project of the University of California, Los Angeles, and the Jet Propulsion 
Laboratory/California Institute of Technology, funded by the National Aeronautics and Space Administration.
This research made use of Montage. It is funded by the National Science Foundation under Grant Number ACI-1440620, and was previously funded by the National Aeronautics 
and Space Administration's Earth Science Technology Office, Computation Technologies Project, under Cooperative Agreement Number NCC5-626 between NASA and the 
California Institute of Technology.

%%%%%%%%%%%%%%%%%%  The Appendices!!!!! %%%%%%%%%%%%%%%%%%%%%%%%%%%%%%%%%%%%%%%%%%%

\appendix

\section{IRAC \& 2MASS Source Selection}\label{irysos}

The removal of contaminating sources and the selection of YSOs in the SMOG field were undertaken following the methods described in 
\citet{gut08} and \citet{gut09} with a number of adjustments to the applied criteria to best suit the character of our dataset, in particular the 
lower background extinction and the greater distances of the SFRs. 
These criteria are described in the following appendix.  

\subsection{Contamination}\label{} 

Candidate AGN contaminants were selected from their location on the  $ [4.5] v. [4.5 - 8.0] $ color-magnitude diagram, with 9287 sources selected. 
\begin{equation}
AGN_1 = [4.5 - 8.0] > 1.2  \mbox{  and  }  [4.5] > 12.5 + ([4.5 - 8.0] - 4.5)/0.2   \mbox{  and  }  [4.5] > 12.5  
\end{equation}
\begin{equation}
\begin{split}
AGN  =  &  [4.5][AGN_1] > 15 - 0.5([4.5 - 8.0][AGN_1] - 1)   \mbox{  and  }  \\
             & ( [4.5][AGN_1] > 14 + 0.5( [4.5 - 8.0][AGN_1] - 2) )  \mbox{  or  }  [4.5][AGN_1] > 15
\end{split}
\end{equation}
AGN share similar colors to YSOs, but are usually fainter, and are removed primarily based on magnitude considerations.  It is possible that fainter 
YSOs also fall into this color-space, however a spatial distribution plot shows that the selected sources are randomly distributed across the field and
a less severe cut would most likely include a higher level of contamination.  

The sources considered to be PAH galaxy contaminants were selected based on their locations on two color-color diagrams and a photometric cut-off at 4.5$\mu$m.   
There were 761 and 802 sources selected as PAH contaminants with the first and second criteria, respectively.  
\begin{eqnarray}
PAH_1 = [4.5 - 5.8] < (2.5/2)([5.8 - 8.0] - 1) \mbox{  and  }  [4.5 - 5.8] < 1.55  \mbox{  and  }  [5.8 - 8.0] > 1 \mbox{  and  }  [4.5] > 11.5  \\
PAH_2 = [3.6 - 5.8] < (3.2/3)(4.5 - 8.0] - 1)  \mbox{  and  }  [3.6 - 5.8] < 2.25  \mbox{  and  }  [4.5 - 8.0] > 1 \mbox{  and  }  [4.5] > 11.5
\end{eqnarray}

Possible knots of shocked emission were identified using IRAC bands 1-3, with 20 sources selected:
\begin{eqnarray}
KNOT = [3.6 - 4.5] > 1.05  \mbox{  and  }  [3.6 - 4.5] > (1.2/0.55)( [4.5 - 5.8] - 0.3) + 0.8   \mbox{  and  }   [4.5 - 5.8] \le 0.85 
\end{eqnarray}

Contamination of the aperture by PAH emission was identified using the following criteria, which found 523 sources.
\begin{eqnarray}
\sigma_{12} = \sqrt{( unc_{3.6}^2 + unc_{4.5}^2 )} \\
\sigma_{23} = \sqrt{( unc_{4.5}^2 + unc_{5.8}^2 )}  \\
PA  =  [3.6 - 4.5] - \sigma_{12} \le 1.5*([4.5 - 5.8] - \sigma_{23} - 1)   \mbox{  and  }  [3.6 - 4.5] - \sigma_{12} \le 0.4
\end{eqnarray}

The SMOG field lies in the direction of the outer galaxy, and thus the extragalactic background is expected to be higher 
since the level of shielding from the galactic center from both the stellar population and dust component is reduced.  
The true level of contaminants remaining in the cleaned catalog is difficult to estimate precisely.  Spectral typing of 
all YSOs would be necessary to confirm their nature.

\subsection{YSO Selection}\label{}

The YSOs were selected from the traditional four color IRAC diagram using the following criteria:
\begin{eqnarray}
\sigma_{12} = \sqrt{( unc_{3.6}^2 + unc_{4.5}^2 )} \\
\sigma_{34} = \sqrt{( unc_{5.8}^2 + unc_{8.0}^2 )}  
\end{eqnarray}
\begin{equation}
\begin{split}
[5.8 - 8.0] \ge 0.3 +  \sigma_{34}    \mbox{  and  }  \\
[5.8 - 8.0]  \le 2 -  \sigma_{34}   \mbox{  and  }   \\
[3.6 - 4.5] \ge 0.2 +  \sigma_{12}   \mbox{  or  }   \\
[5.8 - 8.0]  \le 2.5 -  \sigma_{34}    \mbox{  and  }   \\
[3.6 - 4.5]  \ge  0.5 +  \sigma_{12} 
\end{split}
\end{equation}
A total of 3299 YSOs were selected by this method.    

A further selection based on the four IRAC bands was also applied:
\begin{eqnarray}
\sigma_{13} = \sqrt{( unc_{3.6}^2 + unc_{5.8}^2 )}  \\
\sigma_{24} = \sqrt{( unc_{4.5}^2 + unc_{8.0}^2 )}  
\end{eqnarray}
\begin{equation}
\begin{split}
[3.6 - 5.8]  \ge 0.5 + \sigma_{13}    \mbox{  and  }  \\
[4.5 - 8.0]  \ge 0.5 +  \sigma_{24}     \mbox{  and  }  \\  
[3.6 - 4.5]  \ge 0.15 + \sigma_{12}    \mbox{  and  }  \\  
[3.6 - 5.8] + \sigma_{13}  \le  (0.14/0.04)( [4.5 - 8.0]  -  \sigma_{24} - 0.5) + 0.5.                                                                       
\end{split}
\end{equation}
There were 3239 sources selected by this method.  

For those sources lacking a detection at 8.0$\mu$m, we applied a selection criteria using the shortest three IRAC bands:
\begin{equation}
\begin{split}
[3.6  - 4.5] -  \sigma_{12} > 0.3    \mbox{  and  }  \\  
[4.5 -  5.8] -  \sigma_{23} > 0.3 
\end{split}
\end{equation}
This is a less stringent cut than applied in \citet{gut09} where only protostellar objects were selected via this method.  
A total of 2721 were selected using this criteria.  

For sources lacking photometry in the mid-IR, a combination of 2MASS and IRAC bands were used to select for less extincted objects. 
Two cuts were made, one where a detection in $J$-band was required and one where it was not.  These methods identified 347 and 353 sources, respectively. 
\begin{eqnarray}
\sigma_{k1} = \sqrt{( unc_{K_s}^2 + unc_{3.6}^2 )}  
\end{eqnarray}
\begin{equation}
\begin{split}
[3.6 - 4.5] -  \sigma_{12}  >  0.101    \mbox{  and  }  \\   
[K_s - 3.6]  -  \sigma_{k1}  >  0.5     \mbox{  and  }  \\   
[K_s - 3.6]  - \sigma_{k1}  >  -2.85714 ( [3.6 - 4.5] - \sigma_{12}  - 0.101)  +  1.5.  
\end{split}
\end{equation}
A final selection criteria simply selected for those sources with $K_s$ and 8.0$\mu$m detections which displayed a large color excess, a method that yielded 449 sources.  
\begin{eqnarray}
\sigma_{k4} = \sqrt{( unc_{K_s}^2 + unc_{8.0}^2 )} \\
%[$K_s$ - 8.0]  - \sigma_{k4}  >  3
|K_s - 8.0] - \sigma_{k4} >  3
\end{eqnarray}

\section{MIPS Source Selection}\label{mipsysos}

The YSO selection criteria for objects with MIPS 24$\mu$m photometry were adapted from our previous studies \citep{gut08, winston}:  
\begin{eqnarray}
\sigma_{4m} = \sqrt{( unc_{8.0}^2 + unc_{24}^2 )}  
\end{eqnarray}
\begin{equation}
\begin{split}
[8.0 - 24] \ge 1.0 + \sigma_{4m}    \mbox{  and  }  \\    
[5.8 - 8.0] \ge -0.1 - \sigma_{34}     \mbox{  or  }  \\     
[8.0 - 24]  \ge 0.6 +  \sigma_{4m}     \mbox{  and  }  \\   
[5.8 - 8.0] \ge 0.2 -  \sigma_{34} 
\end{split}
\end{equation}
\begin{equation}
\begin{split}
[8.0 - 24] \ge 1.0 + \sigma_{4m}    \mbox{  and  }  \\    
[3.6 - 4.5] \ge -0.1 - \sigma_{12}     \mbox{  or  }  \\     
[8.0 - 24]  \ge 0.6 +  \sigma_{4m}     \mbox{  and  }  \\   
[3.6 - 4.5] \ge 0.2 -  \sigma_{12} 
\end{split}
\end{equation}
These criteria search for embedded objects with only longer wavelength detections, and those with detections across the IRAC bandpasses.

\section{IRAC \& UKIDSS Source Selection}\label{ukysos}

The UKIDSS near-IR photometry and IRAC source selection criteria were:   
\begin{equation}
\begin{split}
[3.6 - 4.5] - \sigma_{12} >  0.101   \mbox{  and  }  \\    
[K - 3.6] - \sigma_{k1}  >  0.5   \mbox{  and  }  \\    
[K -3.6] - \sigma_{k1}  >  -2.85714*([3.6 - 4.5] - \sigma_{12} - 0.101) + 0.5)  
\end{split}
\end{equation}
\begin{equation}
\begin{split}
[K - 8.0] - \sigma_{k4}  >  2   \mbox{  and  }  \\    
[K]  >  12    \mbox{  and  }  \\   
[K]  >  12 + ( [K - 8.0] - 5 ) 
\end{split}
\end{equation}
These mirror the criteria used with the 2MASS selection but account for the saturation of the brighter sources.

\section{WISE Source Selection}\label{wise_ysos}

Following the process laid out by \citet{fischer}, spurious detections were cleaned from the catalog.  The first step was to remove those sources with flags in 
bands W1, W2, and W3.  Upper limits in bands W1, W2, W3 are then removed, and a saturation cut-off of $W1 > 5$ applied to the data.  
The remaining catalog contained 184,912 sources, or $\sim31.8\%$ of the original catalog.  
The contaminating background galaxies and source selection criteria for the {\it WISE} data were taken from \citet{koenig, fischer} and adapted to the 
requirements of the SMOG field.  
We adjusted the criteria for removal of AGN and star-forming galaxy contaminants slightly from those of \citet{koenig} as follows:
\begin{equation}
\begin{split}
SFG =  & [W2 - W3] > 2.3   \mbox{  and  }  \\
            & [W1 - W2]  < 1.0   \mbox{  and  }  \\
            & [W1 - W2] < 0.46([W2 - W3] - 0.78)   \mbox{  and  }  \\
            & [W1] > 14 
\end{split}
\end{equation}
\begin{equation}
\begin{split}
AGN = & [W1] > 1.8( [W1 - W3] + 4.1 ) \mbox{  and  } \\
            & [W1] > 14  \mbox{  or  }   \\
            & [W1]  >  [W1 - W3] + 10.0  
\end{split}
\end{equation}
Of those 184,912 sources, there were 58,845 star forming galaxies and 167,787 AGN identified as contaminants, leaving a cleaned catalog of 16,689 sources.  

The clean catalog was then searched for YSOs according to criteria taken from the \citet{fischer} and \citet{koenig} papers. 
From these the combined 2MASS and {\it WISE} diagram returned 664 YSOs, there were 90 transition disk candidates and 165 YSOs identified from the four-band 
WISE diagram, and 271 class I and 433 class II sources identified from the {\it WISE} 3-band diagram.

%%%%%%%%%%%%%%%%%%  The Tables!!!!! %%%%%%%%%%%%%%%%%%%%%%%%%%%%%%%%%%%%%%%%%%%

\begin{deluxetable*}{ccl}
\tablecolumns{3}
\tablecaption{SMOG Field YSOs:  Photometry Table Description }
\tablehead{   
\colhead{Column Number} & \colhead{Column ID}  & \colhead{Description}  
}
\startdata
1 &  SF ID &  SMOG Field ID      \\  
2 &  GLIMPSE ID &   GLIMPSE  ID     \\  
3 &  2MASS ID &  2MASS  ID     \\  
4 &  RA &  Right Ascension     \\  
5 &  Dec  &   Declination     \\  
6 &  IRAC det. &  selected by IRAC photometry    \\  
7 &  MIPS det. &  selected by MIPS  photometry   \\  
8 &  UKIRT det. & selected by UKIRT photometry    \\  
9 &  J$_{2M}$ & 2MASS J-band     \\  
10 &  eJ$_{2M}$ &   2MASS J-band uncertainty     \\  
11 &  H$_{2M}$ &  2MASS H-band     \\  
12 &  eH$_{2M}$ &  2MASS H-band uncertainty     \\  
13 &  K$_{2M}$ &  2MASS Ks-band     \\  
14 &  eK$_{2M}$ &  2MASS Ks-band uncertainty     \\  
15 &  J$_{UK}$ & UKIRT J-band     \\  
16 &  eJ$_{UK}$ &  UKIRT J-band uncertainty     \\  
17 &  H$_{UK}$ &  UKIRT H-band     \\  
18 &  eH$_{UK}$ &  UKIRT H-band uncertainty     \\  
19 &  K$_{UK}$ &  UKIRT K-band     \\  
20 &  eK$_{UK}$ &  UKIRT K-band uncertainty     \\  
21 &  3.6 & IRAC band 1     \\  
22 &  e3.6 & IRAC band 1 uncertainty     \\  
23 &  4.5 & IRAC band 2     \\  
24 &  e4.5 & IRAC band 2 uncertainty     \\  
25 &  5.8 & IRAC band 3     \\  
26 &  e5.8 & IRAC band 3 uncertainty     \\  
27 &  8.0 & IRAC band 4     \\  
28 &  e8.0 & IRAC band 4 uncertainty     \\  
29 &  24 & MIPS band 1     \\  
30 &  e24 & MIPS band 1 uncertainty     \\  
31 &  $\alpha_{IRAC}$ & SED slope 3.6-8.0$\mu$m     \\  
32 &  $\alpha_{MIPS}$ &   SED slope 3.6-24$\mu$m     \\  
33 &  Class IRAC & Evolutionary Class: IRAC bands     \\  
34 &  Class MIPS & Evolutionary Class: IRAC \& MIPS-1 bands     \\  
35 &  Class Trans & Evolutionary Class: Transition Disk Candidate     \\  
36 &  Cl \# & Number of Cluster containing YSO     \\  
37 &  sCl \# & Number of Subcluster containing YSO     \\  
\enddata
\label{tab:IRphot} 
\tablecomments{Table \ref{tab:IRphot} is published 
in its entirety in the machine readable format.  The column identifiers are 
shown here for guidance regarding its form and content.}
\end{deluxetable*}

\startlongtable

\begin{deluxetable*}{cchhcchhhhhhhhc}

\tabletypesize{\scriptsize}
\tablecaption{SMOG Field Young Stellar Objects: SIMBAD matches within 1 arcsec  \label{tab:Simbad} }
\tablehead{ 
\colhead{SF} & \colhead{Glimpse}  & \nocolhead{RA}  & \nocolhead{Dec} & \colhead{SIMBAD}   & \colhead{OTYPE}  & \nocolhead{SIMBAD RA} & \nocolhead{SIMBAD Dec}  &\nocolhead{B} & \nocolhead{V} & 
\nocolhead{R} & \nocolhead{J} & \nocolhead{H} & \nocolhead{K$_s$} & \colhead{SP\_TYPE}     \\
\colhead{ID}  & \colhead{ID}  & \nocolhead{(J2000)} & \nocolhead{(J2000)} & \colhead{ID}  & \colhead{}  & \nocolhead{(J2000)}  & \nocolhead{(J2000)} &\nocolhead{mag} & \nocolhead{mag}  & \nocolhead{mag} & 
\nocolhead{mag} & \nocolhead{mag} & \nocolhead{mag} &\colhead{} 
}
\startdata
SRC27 & SSTSMOGA G106.1258+01.0513 & 338.326613 & 59.33885 & IRAS 22314+5904 & Star & 22 33 18.45 & +59 20 20.3 &  &  &  &  &  &  &  \\
SRC52 & SSTSMOGA G107.9079+02.2809 & 340.23463 & 61.28648 & IRAS 22390+6101 & Star & 22 40 56.32 & +61 17 11.4 &  &  &  & 9.325 & 7.549 & 6.495 &  \\
SRC118 & SSTSMOGA G106.0932+01.1394 & 338.184048 & 59.398508 & IRAS 22308+5908 & Star & 22 32 44.24 & +59 23 54.6 &  &  &  &  &  &  &  \\
SRC210 & SSTSMOGA G106.1866+02.3247 & 337.129623 & 60.463586 & V* V675 Cep & Candidate\_LP* & 22 28 31.12 & +60 27 49.0 & 14.8 & 12 & 11.8 & 6.187 & 4.983 & 4.527 &  \\
SRC217 & SSTSMOGA G108.5792+02.5484 & 341.208652 & 61.83991 & IRAS 22429+6134 & Star & 22 44 50.07 & +61 50 23.4 &  &  &  &  &  &  &  \\
SRC254 & SSTSMOGA G108.5944+00.4895 & 343.161178 & 60.012353 & 2MASS J22523871+6000445 & IR & 22 52 38.71 & +60 00 44.5 &  &  &  &  & 13.449 & 10.95 &  \\
SRC561 & SSTSMOGA G106.5735+00.5228 & 339.59871 & 59.102359 & [D75b] Em* 22-073 & Em* & 22 38 23.70 & +59 06 08.5 & 14 & 12.5 & 13.1 & 11.29 & 10.944 & 10.666 &  \\
SRC603 & SSTSMOGA G103.4831+01.9991 & 333.007481 & 58.715673 & IRAS 22103+5828 & IR & 22 12 01.754 & +58 42 56.20 &  &  &  & 12.61 & 10.91 & 9.8 &  \\
SRC618 & SSTSMOGA G103.4890+00.5787 & 334.527043 & 57.543425 & EM* GGR   56 & Em* & 22 18 06.519 & +57 32 36.40 & 12.22 & 11.42 &  & 9.151 & 8.766 & 8.363 & B \\
SRC707 & SSTSMOGA G102.4061+00.8939 & 332.539004 & 57.192873 & LS III +56   27 & Star & 22 10 09.29 & +57 11 34.6 & 13.4 &  & 12.2 & 10.935 & 10.567 & 10.372 & OB- \\
\enddata
\tablecomments{Table \ref{tab:Simbad} is published 
in its entirety in the machine readable format.  A portion is
shown here for guidance regarding its form and content.}
\end{deluxetable*}

\startlongtable
\begin{deluxetable*}{ccccccccch}
\tablecolumns{10}
\tabletypesize{\scriptsize}
\tablecaption{SMOG Field Young Stellar Objects: Cluster Descriptions }
\tablehead{ 
\colhead{Cluster} & \colhead{Cluster} & \colhead{YSO}  & \colhead{Subcluster}  & \colhead{Central} & \colhead{Central}   & \colhead{Circular}  & \colhead{Branch} & \colhead{WISE HII}  &\nocolhead{SIMBAD}    \\
\colhead{ID} & \colhead{Name}  & \colhead{\#}  & \colhead{RA} & \colhead{Dec}   & \colhead{Radius}  & \colhead{Length} & \colhead{ID}  &\nocolhead{IDs}    \\
\colhead{}  & \colhead{} & \colhead{}  & \colhead{} & \colhead{(J2000)} & \colhead{(J2000)}  & \colhead{deg}  & \colhead{deg}  & \colhead{} &\nocolhead{} 
}
\startdata
0 & 'G106.48+1.0' & 155 & 5 & 339.03101153 & 59.4870041152 & 0.429 & 0.015 & ['G106.241+00.957', 'G106.499+00.925'] & ['HD 214418', 'IRAS 22344+5909', 'GT 2234+593', 'IRAS 22324+5904', 'IRAS 22327+5903', 'IRAS 22340+5903', 'IRAS 22345+5917', 'GT 2232+590', 'GB6 B2234+5922', '[YDM97] CO  24', 'GB6 B2232+5904', 'NVSS J223437+593011', 'NVSS J223634+593212', 'TGU H658', 'TYC 3995-510-1', 'TYC 3995-556-1', 'TYC 3995-614-1', 'DOBASHI 3319'] \\
1 & 'G106.33+0.04' & 40 & 3 & 339.642358587 & 58.558101249 & 0.123 & 0.013 & [] & ['GT 2237+583', 'TYC 3995-671-1', 'TYC 3995-1313-1', '[D75b] Em* 22-075', 'TYC 3995-591-1', 'TYC 3995-1118-1', 'TYC 3995-1182-1'] \\
2 & 'G108.57+0.3' & 321 & 13 & 343.177203295 & 59.8416664739 & 0.809 & 0.028 & ['G108.603+00.494'] & ['HD 240079', 'V* CV Cep', 'PN K  3-85', 'IRAS 22488+5914', 'GT 2248+591', 'IRAS 22491+5915', 'GPSR 108.072+0.087', 'BD+58  2488', 'TYC 3996-72-1', 'TYC 3996-1148-1', 'NVSS J225021+592508', 'TYC 3996-590-1', 'HD 240089', 'HR  8707', 'CSI+59-22518', 'CSI+59-22534', 'HD 216573', 'GT 2249+592', 'IRAS 22506+5944', '2MASX J22544960+5952481', 'IRAS 22519+5947', 'GT 2250+592', 'GT 2250+596', 'GT 2251+595', 'IRAS 22489+5927', 'IRAS 22491+5921', 'IRAS 22493+5945', 'IRAS 22500+5914', 'IRAS 22502+5919', 'IRAS 22502+5944', 'IRAS 22504+5929', 'IRAS 22520+5932', 'IRAS 22522+5929', 'IRAS 22526+5920', '2MASS J22524957+5957075', 'GT 2251+598', 'GPSR 108.298+0.044', 'GPSR 108.300+0.040', 'GPSR 108.356+0.377', 'GPSR 108.360+0.378', 'GPSR 108.569+0.466', 'GPSR 108.759+0.488', 'GPSR 108.779+0.393', 'GEN\# +6.42001097', 'GEN\# +6.42001106', 'RRF 1043', 'TYC 3996-240-1', 'TYC 3996-250-1', 'TYC 3996-332-1', 'TYC 3996-384-1', 'TYC 3996-490-1', 'TYC 3997-233-1', 'TYC 3997-609-1', 'TYC 3997-691-1', 'TYC 3997-695-1', 'TYC 3997-973-1', 'TYC 3997-1316-1', 'TYC 3997-1694-1', 'TYC 4278-1003-1', '[HLB98] Onsala 170', '[R2003] 150', '[CF95] 56', '[PBC91] B225038.7+594450', '2MASS J22523871+6000445', 'NVSS J225120+594824', 'GB6 B2249+5914', 'NVSS J225206+592849', 'NVSS J225349+600500', 'TGU H699 P14', 'JCMTSE J225238.5+600044', 'JCMTSF J225447.0+595232', 'JCMTSF J225447.6+595339', 'JCMTSF J225450.7+595421', 'JCMTSF J225357.9+600223', 'TYC 3997-245-1', 'TYC 3996-424-1', 'TYC 3996-438-1', 'TYC 3996-520-1', 'TYC 3996-564-1', 'TYC 3996-578-1', 'TYC 3996-644-1', 'TYC 3996-646-1', 'TYC 3996-692-1', 'TYC 3996-708-1', 'TYC 3996-1062-1', 'TYC 3996-1164-1', 'TYC 3997-147-1', 'TYC 3997-309-1', 'TYC 3997-1142-1', 'TYC 3997-1258-1', 'TYC 4278-1152-1', 'DOBASHI 3360', 'DOBASHI 3362', 'DOBASHI 3363', 'DOBASHI 3364', 'DOBASHI 3365', 'DOBASHI 3369', 'DOBASHI 3370', 'DOBASHI 3373', 'DOBASHI 3379', 'DOBASHI 3384', 'MSX6C G108.5955+00.4935', '[FMT2009] 151'] \\
3 & 'G106.08+1.0' & 11 & 0 & 338.307555636 & 59.2724963636 & 0.089 & 0.031 & [] & [] \\
4 & 'G106.41+3.1' & 27 & 1 & 336.675116071 & 61.2335122011 & 0.163 & 0.034 & [] & ['LDN 1195', 'LDN 1196', '[LM99] L1195', 'EM* GGR   71', 'GN 22.24.9', 'DOBASHI 3317'] \\
5 & 'G106.36+0.45' & 8 & 0 & 339.308996875 & 58.936327375 & 0.073 & 0.031 & [] & ['LDN 1197', '[LM99] L1197', 'TGU H655', 'JCMTSE J223702.0+585735', 'DOBASHI 3316'] \\
6 & 'G106.18+0.14' & 30 & 2 & 339.30691536 & 58.576359 & 0.183 & 0.032 & ['G106.142+00.129'] & ['HD 214541', 'IRAS 22350+5817', 'TYC 3995-489-1', 'HBHA 5705-50'] \\
7 & 'G108.51+2.44' & 8 & 0 & 341.185721 & 61.7164695 & 0.093 & 0.047 & [] & ['IRAS 22429+6134', 'TYC 4265-198-1', 'DOBASHI 3367'] \\
8 & 'G107.97+0.49' & 11 & 1 & 341.9835166 & 59.7509637 & 0.130 & 0.077 & [] & ['TYC 3996-110-1'] \\
9 & 'G108.69+1.63' & 17 & 1 & 342.296550864 & 61.0738400643 & 0.172 & 0.039 & [] & ['TYC 4265-2-1', 'TYC 4265-549-1'] \\
\enddata
\label{tab:clusters} 
\tablecomments{ Table \ref{tab:clusters} is published 
in its entirety in the machine readable format.  A portion is
shown here for guidance regarding its form and content.}

\end{deluxetable*}

\begin{deluxetable*}{ccl}
\tablecolumns{3}
\tablecaption{SMOG Field YSOs: WISE Photometry Table Description  }
\tablehead{   
\colhead{Column Number} & \colhead{Column ID}  & \colhead{Description}  
}

\startdata
1 &  WISE ID &  WISE identifier     \\  
2 &  SMOG ID &   SMOG   identifier     \\  
3 &  RA &  Right Ascension     \\  
4 &  Dec  &   Declination     \\  
5 &  J$_2m$ & 2MASS J-band     \\  
6 &  eJ$_2m$ &   2MASS J-band uncertainty     \\  
7 &  H$_2m$ &  2MASS H-band     \\  
8 &  eH$_2m$ &  2MASS H-band uncertainty     \\  
9 &  K$_2m$ &  2MASS Ks-band     \\  
10 &  eK$_2m$ &  2MASS Ks-band uncertainty     \\  
11 &  3.5 & WISE band 1     \\  
12 &  e3.5 & WISE band 1 uncertainty     \\  
13 &  4.6 & WISE band 2     \\  
14 &  e4.6 & WISE band 2 uncertainty     \\  
15 &  12 & WISE band 3     \\  
16 &  e12 & WISE band 3 uncertainty     \\  
17 &  22 & WISE band 4     \\  
18 &  e22 & WISE band 4 uncertainty     \\  
19 &  Class & Evolutionary classification     \\  
20 &  IRAC & Also detected as an IRAC source    \\  
\enddata
\label{tab:WISEphot}
\tablecomments{Table \ref{tab:WISEphot} is published 
in its entirety in the machine readable format.  The column identifiers are 
shown here for guidance regarding its form and content.}
\end{deluxetable*}

\startlongtable
\begin{deluxetable*}{ccchchhchhchhchhchhchhchhchhchh}
\tabletypesize{\scriptsize}
\tablecaption{SMOG Field Young Stellar Objects: SED Fitter Results }
\tablehead{ 
\colhead{SF} & \colhead{Cl}  & \colhead{Dist}  & \nocolhead{N$_data$} & \colhead{$\chi^{2}$}   & \nocolhead{$\chi^{2}_{l}$}  & \nocolhead{$\chi^{2}_{h}$} & \colhead{A$_V$}  &\nocolhead{A$_{V,l}$} & \nocolhead{A$_{V,h}$} & 
\colhead{M$_*$} & \nocolhead{M$_{*,l}$} & \nocolhead{M$_{*,h}$} & \colhead{Age} &\nocolhead{Age$_l$} & \nocolhead{Age$_h$} & \colhead{M$_{disk}$} & \nocolhead{M$_{disk, l}$} & \nocolhead{M$_{disk, h}$} & 
\colhead{M$_{envl}$} & \nocolhead{M$_{envl, l}$} & \nocolhead{M$_{envl, h}$} & \colhead{$\dot{M}$}  & \nocolhead{$\dot{M}_l$} & \nocolhead{$\dot{M}_h$} & \colhead{T$_{*}$} & \nocolhead{T$_{*,l}$} & \nocolhead{T$_{*,h}$} & 
\colhead{L$_{*}$} & \nocolhead{L$_{*,l}$} & \nocolhead{L$_{*,h}$}      \\
\colhead{ID}  & \colhead{\#}  & \colhead{kpc} & \nocolhead{} & \colhead{}  & \nocolhead{}  & \nocolhead{}  & \colhead{mag} &\nocolhead{mag} & \nocolhead{mag}  & \colhead{M$_{\odot}$} & \nocolhead{M$_{\odot}$} & \nocolhead{M$_{\odot}$} & 
\colhead{yr} &\nocolhead{yr} & \nocolhead{yr} & \colhead{M$_{\odot}$} & \nocolhead{M$_{\odot}$} & \nocolhead{M$_{\odot$}} & \colhead{M$_{\odot}$} & \nocolhead{M$_{\odot}$} & \nocolhead{M$_{\odot}$} & \colhead{M$_{\odot}$/yr} & \nocolhead{M$_{\odot}$/yr} & \nocolhead{M$_{\odot}$/yr} & \colhead{K} & \nocolhead{K} & \nocolhead{K} & 
\colhead{L$_{\odot}$} & \nocolhead{L$_{\odot}$} & \nocolhead{L${\odot}$} 
}
\startdata
SRC6 & 0 & 6.0 & 4 & 0.073 & 0.073 & 12.070 & 36.580 & 0.000 & 40.000 & 2.596 & 0.117 & 8.418 & 2.766e+06 & 2.888e+03 & 9.268e+06 & 8.765e-08 & 1.841e-08 & 3.614e-01 & 3.217e+00 & 1.430e+00 & 1.448e+01 & 0.000e+00 & 0.000e+00 & 3.242e-03 & 5.752e+03 & 2.666e+03 & 1.303e+04 & 1.958e+01 & 8.959e-01 & 1.298e+03 \\
SRC16 & 0 & 6.0 & 5 & 0.613 & 0.613 & 15.150 & 32.110 & 0.000 & 40.000 & 1.993 & 0.166 & 2.539 & 4.062e+05 & 1.072e+03 & 4.073e+05 & 1.754e-02 & 6.251e-05 & 8.794e-02 & 3.725e+00 & 1.958e+00 & 1.667e+01 & 4.882e-06 & 3.781e-07 & 2.014e-05 & 4.583e+03 & 2.900e+03 & 4.674e+03 & 1.260e+01 & 1.884e+00 & 3.072e+01 \\
SRC60 & 0 & 6.0 & 4 & 2.006 & 2.006 & 13.940 & 35.130 & 0.000 & 40.000 & 1.443 & 0.102 & 10.780 & 4.462e+05 & 1.098e+03 & 9.715e+06 & 2.603e-02 & 7.584e-06 & 2.593e-01 & 4.149e+00 & 1.950e+00 & 2.390e+01 & 1.073e-07 & 0.000e+00 & 1.718e-03 & 4.447e+03 & 2.541e+03 & 1.132e+04 & 6.725e+00 & 9.817e-01 & 2.269e+03 \\
SRC66 & 0 & 6.0 & 5 & 2.687 & 2.687 & 17.650 & 19.690 & 0.000 & 40.000 & 0.745 & 0.162 & 4.056 & 1.965e+05 & 1.365e+03 & 5.853e+06 & 6.353e-03 & 1.165e-07 & 7.641e-02 & 9.685e+00 & 2.298e+00 & 2.390e+01 & 4.222e-06 & 0.000e+00 & 1.431e-04 & 4.007e+03 & 2.893e+03 & 1.444e+04 & 5.877e+00 & 1.865e+00 & 2.174e+02 \\
SRC74 & 0 & 6.0 & 4 & 0.139 & 0.139 & 12.140 & 22.330 & 0.000 & 40.000 & 2.993 & 0.284 & 7.368 & 1.273e+06 & 5.241e+03 & 9.000e+06 & 6.148e-04 & 2.531e-08 & 3.922e-01 & 1.559e+00 & 1.559e+00 & 1.158e+01 & 0.000e+00 & 0.000e+00 & 2.964e-03 & 5.125e+03 & 3.319e+03 & 9.742e+03 & 1.383e+01 & 2.644e+00 & 6.502e+02 \\
SRC79 & 0 & 6.0 & 5 & 0.656 & 0.656 & 15.620 & 18.900 & 0.000 & 34.670 & 1.523 & 0.515 & 3.595 & 3.219e+05 & 1.408e+05 & 9.713e+06 & 6.069e-05 & 3.593e-09 & 1.050e-01 & 4.805e+00 & 1.815e+00 & 8.324e+00 & 4.153e-07 & 0.000e+00 & 1.295e-05 & 4.425e+03 & 3.789e+03 & 1.326e+04 & 9.841e+00 & 2.675e+00 & 1.363e+02 \\
SRC91 & 0 & 6.0 & 3 & 0.005 & 0.005 & 9.004 & 33.580 & 0.000 & 40.000 & 2.086 & 0.107 & 13.870 & 5.322e+05 & 1.154e+03 & 9.891e+06 & 9.047e-04 & 0.000e+00 & 8.822e-01 & 3.475e+00 & -1.000e+00 & 2.790e+01 & 1.629e-06 & 0.000e+00 & 5.122e-03 & 4.646e+03 & 2.585e+03 & 1.721e+04 & 9.658e+00 & 1.104e+00 & 4.464e+03 \\
SRC93 & 0 & 6.0 & 5 & 2.776 & 2.776 & 17.740 & 4.278 & 0.000 & 17.220 & 2.030 & 0.130 & 6.750 & 4.709e+04 & 1.283e+03 & 9.713e+06 & 1.068e-03 & 4.850e-08 & 2.716e-01 & 8.707e+00 & 1.663e+00 & 2.058e+01 & 1.673e-05 & 0.000e+00 & 5.331e-04 & 4.312e+03 & 2.738e+03 & 1.385e+04 & 4.010e+01 & 1.037e+00 & 3.859e+02 \\
SRC99 & 0 & 6.0 & 4 & 0.287 & 0.287 & 12.280 & 8.046 & 0.000 & 40.000 & 0.414 & 0.104 & 8.418 & 1.604e+05 & 2.407e+03 & 9.891e+06 & 1.781e-03 & 5.931e-09 & 6.401e-01 & 5.948e+00 & 1.174e+00 & 2.361e+01 & 3.171e-05 & 0.000e+00 & 3.242e-03 & 3.625e+03 & 2.616e+03 & 1.303e+04 & 1.906e+00 & 6.167e-01 & 1.298e+03 \\
SRC100 & 0 & 6.0 & 5 & 2.512 & 2.512 & 17.470 & 39.450 & 13.490 & 40.000 & 2.803 & 0.630 & 3.949 & 1.983e+06 & 1.343e+05 & 9.841e+06 & 2.536e-03 & 6.997e-07 & 1.050e-01 & 4.349e+00 & 1.830e+00 & 8.622e+00 & 0.000e+00 & 0.000e+00 & 1.862e-05 & 5.436e+03 & 3.921e+03 & 1.220e+04 & 1.717e+01 & 3.948e+00 & 8.277e+01 \\
\enddata
\label{tab:SEDfit}
\tablecomments{Table \ref{tab:SEDfit} is published in its entirety in the machine readable format.  A portion is shown here for guidance regarding its form and content. {\bf A number of columns, such as the upper and lower limits, are not displayed but are included in the full, online version of the table. }  }
\end{deluxetable*}

\begin{deluxetable*}{crrrrr}
\tablecaption{WISE and IRAC source matching \label{wise_irac_match} }
\tablewidth{0pt}
\tablehead{
\colhead{WISE} & \colhead{}  & \colhead{Matching} & \colhead{Full} & \colhead{Clean} & \colhead{IRAC}  \\
\colhead{Catalog} & \colhead{Size}  & \colhead{Radius} & \colhead{IRAC} & \colhead{IRAC} & \colhead{YSOs} 
}
\startdata
Full   &  581705 &  1.0 &  460509  &  456530   &  2843   \\
Full   &  581705 &  1.5 &  514063  &  509518   &  3115   \\
Clean   &  16689 &  1.0 &    &  14811   &  913   \\
Clean   &  16689 &  1.5 &    &  15199   &  966   \\
YSOs   &  931 &  1.0 &  711  &  700   & 634   \\
YSOs   &  931 &  1.5 &  770  &  759  &  672   \\
\enddata
\end{deluxetable*}

\begin{deluxetable*}{rrrrrrrrrrrrr}
\tablecaption{WISE HII distances to matched clusters \label{tab:wise_distances} }
\tablewidth{0pt}
\tablehead{
\colhead{Cluster} & \colhead{\#0}  & \colhead{\#2} & \colhead{\#10} & \colhead{\#11} & \colhead{\#25} & \colhead{\#30} & 
\colhead{\#32}  & \colhead{\#39} & \colhead{\#41} & \colhead{\#60} & \colhead{\#64} & \colhead{\#66}
}
\startdata
 Distance (kpc) & 6.0 & 2.5 & 5.7 & 2.7 & 7.8 & 3.1 & 7.4 & 2.6 & 7.3 & 7.2 & 4.4  & 5.5  \\ 
\enddata
\end{deluxetable*}

%%%%%%%%%%%%%%%%%%  The Figures!!!!! %%%%%%%%%%%%%%%%%%%%%%%%%%%%%%%%%%%%%%%%%%%

%%%%%%%%%%%%%%%%%%  The Bibliography!!!!! %%%%%%%%%%%%%%%%%%%%%%%%%%%%%%%%%%%%%%%%%%%

%%%%%%%%%%%%%%%%%%%%  END OF PAPER!!!!!!!  %%%%%%%%%%%%%%%%%%%%%%%%%%%%%%%%%%%%%%%%%%

\end{document}